\documentclass[12pt, a4paper, twoside]{article}
\pdfoutput=1
\usepackage[a4paper, left=3cm,right=2cm]{geometry}
\usepackage{hyperref}
\usepackage{amsfonts}
\usepackage{amsmath}
\usepackage{empheq}
\usepackage{setspace}
\usepackage{color}

\usepackage{cite}

\usepackage{graphicx}
\usepackage[font=small, font+=it, format=hang, margin={1cm, 1cm}]{caption}
\graphicspath{{figure/}}

\usepackage{dcolumn}
\usepackage{bm}

\newcommand{\bea}{\begin{eqnarray}}
\newcommand{\eea}{\end{eqnarray}}
\newcommand{\be}{\begin{equation}}
\newcommand{\ee}{\end{equation}}
\def\nn{\nonumber}

\def\p{\partial}
\def\eps{\epsilon}
   
\title{Near-horizon Extreme Kerr Magnetospheres}

\begin{document}


\begin{titlepage}

\begin{flushright}\vspace{-3cm}
{\small
\today }\end{flushright}
\vspace{0.5cm}

\begin{center}

\centerline{{\Large{\bf{Near-horizon Extreme Kerr Magnetospheres}}}} \vspace{10mm}

\centerline{\large{\bf{G. Comp\`{e}re\footnote{e-mail: gcompere@ulb.ac.be}, R. Oliveri\footnote{e-mail:
roliveri@ulb.ac.be}}}}

\vspace{5mm}
\normalsize
\bigskip\medskip
\textit{Universit\'{e} Libre de Bruxelles and International Solvay Institutes\\
CP 231, B-1050 Brussels, Belgium
}

\vspace{5mm}

\begin{abstract}
\noindent 
{Analytical solutions to force-free electrodynamics around black holes are fundamental for building simple models of accretion disk and jet dynamics. 
We present a (nonexhaustive) classification of complex highest-weight solutions to the force-free equations in the near-horizon region of the extremal Kerr black hole. 
Several classes of real magnetically dominated or null solutions, either axisymmetric or nonaxisymmetric, are described which admit finite energy and angular momentum with respect to the asymptotically flat observer. Subtleties related to the velocity of light surface in the near-horizon region are discussed.
}

\end{abstract}


\end{center}

\end{titlepage}

\newpage
\tableofcontents

\section{Introduction and outline}

Black holes surrounded by an accreting plasma admit a rich dynamics which is captured by the framework of general relativistic magnetohydrodynamics. 
Under the assumptions that the plasma does not feel the Lorentz force and that the matter backreaction is negligible, the dynamics reduces to the so-called force-free electrodynamics in the fixed background geometry of the black hole \cite{Blandford:1977ds}. This approximation has been numerically shown to be valid close to the poles in certain regimes \cite{2004ApJ...611..977M,2007MNRAS.375..513M}. The force-free equations are still nonlinear and therefore only few exact analytical solutions are known; see \cite{1973ApJ...180L.133M,1976MNRAS.176..465B,Blandford:1977ds,2007GReGr..39..785M,2011PhRvD..83l4035L,Menon:2011zu,Brennan:2013kea,Lupsasca:2014pfa,Gralla:2015vta} for a nearly exhaustive list. In this paper we will continue the recent ongoing effort in deriving new analytical solutions which could be used as simple models for accretion around rotating black holes. 

Recent observations indicate that near-extremal spinning black holes exist in nature \cite{Blum:2009ez,2013ApJ...775L..45M,Gou:2013dna,Brenneman:2006hw,2009Natur.459..540F}. For such black holes, a throat geometry forms in the vicinity of the horizon. This geometry is the so-called near-horizon extremal Kerr region, which admits enhanced $SL(2,\mathbb R)$ symmetry \cite{Bardeen:1999px}. 
Such symmetries are familiar in conformal field theories where fields typically fall into highest-weight representations of the conformal group. Building highest-weight representations in terms of the electromagnetic gauge field can therefore be promoted as a solution generating technique for force-free electrodynamics in the vicinity of near-extremal black holes \cite{Lupsasca:2014pfa}. 

A caveat is that not all the regions of the near-horizon extremal Kerr black hole are physical. Indeed, the nonextremal Kerr black hole contains a velocity of light surface, defined as the codimension one surface away from the horizon where the Killing horizon generator becomes null, strictly outside of the horizon. The fact that it asymptotes at the equator to the horizon in the extremal limit leads to the unphysical presence of a velocity of light surface in the near-horizon limit bounded by two polar angles that are symmetric with respect to the equator. As a consequence, no global timelike Killing vector exists in the near-horizon region and no quantum field vacuum can be defined \cite{Kay:1988mu,Ottewill:2000qh,Ottewill:2000yr} (this was also emphasized in \cite{Amsel:2009ev}). So it would not make sense to define a $e^+-e^-$ plasma in that region. 
This unphysical feature is not present away from the extremal limit. We take therefore the point of view that one should build physical solutions in the region between the north pole and the northern velocity of light surface only (and similarly in the south region). We will call the union of these north and south regions away from the velocity of light surfaces the physical near-horizon region.

Several classes of solutions have been found using the highest-weight technique \cite{Lupsasca:2014pfa,Lupsasca:2014hua,Zhang:2014pla}. As in each solution generating technique, one distinguishes between the set of generated formal solutions and the set of physical solutions. First, highest-weight solutions are generally complex, and only when the current is proportional to its complex conjugate up to an arbitrary function could one superpose the real and imaginary parts of the solution to get a real solution, as shown in \cite{Lupsasca:2014hua}. Second, many real solutions are electrically dominated in the physical near-horizon region. As explained e.g. in \cite{Komissarov:2004ms}, electrically dominated force-free solutions are unphysical since there exists a local inertial frame for which charged matter moves at a drift velocity higher than the speed of light. Null solutions (with electric and magnetic fields of equal magnitude) correspond to the borderline case where the charged matter moves at the speed of light. We will encounter all such types of solutions but we will seek for magnetically dominated or null solutions. 

Moreover, one should always require that solutions have finite energy and angular momentum which in the case of highest-weight solutions imposes a bound on their weight. Here, there are two possible notions of energy. Either one insists on having finite energy in the near-horizon region, which might be useful for discussing holography in near-horizon geometries, or one insists on having a finite and nonvanishing energy with respect to the asymptotically flat observer, which we consider as the physical requirement. The latter requirement imposes bounds on the weight. For such weights, the variational principle should be well-defined in the near-horizon region. 

Our main result is the final list of near-horizon solutions which admit finite and nonvanishing energy and angular momentum with respect to the asymptotically flat observer which we could build from the complex highest-weight solutions. We limited ourselves to near-horizon solutions which led to computable energy or angular momentum flux in the asymptotically flat region.  While we found several classes of axisymmetric or non-axisymmetric, magnetically dominated or null solutions, we could not conclude whether or not these solutions should be considered physical. Indeed, they are singular at the horizon, except one notable null solution found in \cite{Lupsasca:2014hua}, and in some cases admit a logarithmic divergence at the velocity of light surface. One can argue that the velocity of light surface of the near-horizon region should have order one corrections upon gluing the asymptotically flat region. If this is the case, this logarithmic divergence might be unphysical. This issue could only be totally settled by considering the extension of the near-horizon solutions in the asymptotically flat geometry, which is a difficult analytical problem. 

Our paper is organized as follows. After a brief review of the near-horizon extremal Kerr geometry, we first present an extension of the formalism of Euler potentials \cite{1997PhRvE..56.2198U} to describe nonaxisymmetric solutions in a canonical form in Sec. \ref{cep} which will be instrumental in expressing the solutions. We then discuss the $SL(2,\mathbb R)$ invariant solutions in detail in order to get some intuition. Our main analysis takes place in Sec. \ref{hwsol}. First, we define a $SL(2,\mathbb R)$ covariant basis of $1$- and $2$-forms to express all the physical quantities in a covariant fashion. We then construct the highest-weight representations of the vector potential, field strength and current.
Second, in Sec. \ref{sec:E}, we analyse the energy and angular momentum fluxes with respect to both an observer in the near-horizon region and an asymptotically flat observer. We are interested in solutions with finite and nonvanishing asymptotically flat energy and angular momentum that are determined from the near-horizon region. This analysis, according to \cite{Gralla:2016jfc}, provides bounds on the weights of the solutions.
Third, we show that in order to obtain highest-weight representations obeying the force-free equations we only need to solve three coupled nonlinear ordinary differential equations. We classify the types of solutions using several criteria (electromagnetic type, admitting or not admitting descendants, etc) and partially reduce the equations to a list of independent solutions. We however identify one class of two coupled nonlinear equations that we were not able to reduce, so our classification, even though it extends the existing literature, is not complete.We then list all the independent complex highest-weight solutions which we obtained.  In Sec. \ref{physSol}, we first build real solutions, from the list  of complex solutions, which are magnetically dominated or null. This preliminary result leads to a list of near-horizon solutions in Sec. \ref{sec:near}. By further imposing the bounds derived in Sec. \ref{sec:E} to have finite and nonvanishing energy and angular momentum which respect to the  asymptotically flat frame, we get the final list of potentially physical near-horizon solutions in Sec. \ref{listphys}. The highest-weight solutions are expressed in terms of five distinct functions of the polar coordinate. Out of these five functions, three solve linear ordinary differential equations (ODEs) and two solve nonlinear ODEs with boundary conditions fixed by regularity requirements as we discuss in Appendix \ref{app:ODE}. We numerically solved explicitly all three linear ODEs. For completeness and future reference, the details of all highest-weight solutions found are listed in Appendix \ref{ID}.

{\bf Note added in v3} After the publication of \cite{Gralla:2016jfc}, we noted two algebraic mistakes in Eq. (97) and (101) in v2 which we corrected in this version, in Eq. \eqref{EE} and \eqref{Eo}. These corrections modify the admissible weights and the list of potentially physical solutions. Moreover, complete bounds on the weights require the analysis of angular momentum as well, as done in \cite{Gralla:2016jfc}. Sec. \ref{sec:E} and \ref{listphys} have been rewritten and the introduction and conclusion have been adapted accordingly. We also commented further upon regularity at the future Poincar\'e horizon.

\section{Near-horizon extremal Kerr geometry}

The near-horizon limit of the near-extremal Kerr black hole has been explicitly written in many references and we refer the reader to those \cite{Bardeen:1999px,Guica:2008mu,Bredberg:2011hp,Compere:2012jk}. It is important to emphasize that depending on how the near-extremality parameter $J-M^2$ is scaled upon taking the near-horizon limit, two distinct coordinate systems might result in the near-horizon extremal Kerr geometry, namely Poincar\'e and black hole coordinates. Such coordinate systems do not cover the entire near-horizon spacetime and can be extended into global coordinates. It is therefore of interest to write a formalism which does not depend upon the choice of coordinate system and which can be easily specialized for each case. 

The four-dimensional near-horizon extremal Kerr metric can be written in a way that makes its $SL(2,\mathbb R) \times U(1)$ symmetries manifest but without choosing a particular coordinate system as
\begin{equation}\label{NHEK}
ds^2 = \Gamma(\theta) \left[  ds_{AdS_2}^2 + d\theta^2 +\gamma^2(\theta) \left[ d\Psi + k \Theta \right]^2 \right].
\end{equation}
Here $\Psi \sim \Psi + 2\pi$ is the azimuthal angle, $\theta \in [0,\pi]$ the polar angle, $ds_{AdS_2}^2$ is the unit metric on two-dimensional anti-de Sitter spacetime $AdS_2$ and $\Theta$ is a left-invariant one-form on $AdS_2$ with norm $-1$. The geometry depends upon the two functions $\gamma,\Gamma$ of the polar angle and upon the constant $k$. We will keep $k$ explicitly here in order to more easily allow for future generalizations to other near-horizon geometries such as e.g. the one coming from the Kerr-Newman black hole which has the form \eqref{NHEK} with $k \neq 1$. For the extremal Kerr near-horizon geometry, we have
\bea
k=1,\qquad \Gamma(\theta) = J \frac{1+\cos^2\theta}{2} ,\qquad \gamma(\theta) = \frac{2\sin\theta}{1+\cos^2\theta}\label{paramKerr}
\eea
where $J$ is the angular momentum. The velocity of light surface is located in the range $[\theta_*,\pi - \theta_*]$ where $\theta_*=\arcsin(\sqrt{3}-1)$. We will mostly consider the physical near-horizon region defined as $[0,\theta_*] \cup [\pi-\theta_*,\pi]$.

We denote the symmetry generators of $SL(2,\mathbb R)$ as $H_0,H_\pm$ and the $U(1)$ symmetry generator as $Q_0 \equiv \p_\Psi$.  For future use, we also define the rescaled generator
$
\hat Q_0 \equiv \frac{Q_0}{Q_0 \cdot Q_0}
$.  When considered as a form after using the metric, it has the convenient expression $\hat Q_0 = d\Psi + k \Theta$ (we denote forms and vectors by the same symbol and they are distinguished by the context).

The generators obey the $SL(2,\mathbb R)\times U(1)$ commutation relations
\begin{equation}\label{algebra}
[ H_0,H_\pm ] = \mp H_\pm,\qquad [H_+,H_-] = 2H_0,\qquad
[ Q_0,H_\pm ] = 0,\qquad [Q_0,H_0] = 0.
\end{equation}

A physical feature of this geometry is the absence of globally timelike Killing vector. Here, we complement the comments given in the Introduction. For a nonextremal black hole in comoving coordinates the Killing horizon generator is null at the horizon and timelike just outside the horizon and it becomes null away from the horizon at the velocity of light surface beyond which it becomes spacelike. At extremality and at the equator the velocity of light surface asymptotes towards the horizon and therefore upon taking the near-horizon limit, the Killing generator might not remain timelike everywhere. The region where the Killing generator remains timelike represents a region where the physics is clearly related to the asymptotically flat region. On the contrary, the region where it becomes spacelike is rather a special feature of the near-horizon extremal limit which disappears in the asymptotically flat region.

For further use, we define the highest-weight scalar $\Phi_{(h,q)}$ of weight $h$ and charge $q$ as 
\begin{equation}
H_+ \Phi_{(h,q)} = 0, \qquad H_0 \Phi_{(h,q)} = h \Phi_{(h,q)},\qquad Q_0 \Phi_{(h,q)} = i q \Phi_{(h,q)},\qquad \partial_\theta \Phi = 0. \label{scalHW}
\end{equation}
We denote
\begin{equation}
\Phi \equiv \Phi_{(1,0)} , \qquad \lambda \equiv \Phi_{(0,1)}  
\end{equation}
Since the definition is linear it readily follows that 
\begin{equation}
\Phi_{(h,q)} =\Phi^h \lambda ^q\, . \label{solHW}
\end{equation}

We will now make contact in the upcoming sections with the three special coordinate systems: Poincar\'e coordinates, global coordinates and black hole coordinates.

\subsection{Poincar\'e coordinates}

We denote Poincar\'e coordinates as $(t,r,\theta,\phi)$. We have $\Psi = \phi$. The metric is 
\bea
ds^2 = \Gamma \left[  \frac{dr^2}{r^2} + d\theta^2 - r^2 dt^2+\gamma^2 \left[ d\phi + k rdt \right]^2 \right].
\eea
We define
\bea
H_+ &=& \sqrt{2}\p_t,\\
H_0&=& t \p_t - r \p_r,\\
H_- &=& \sqrt{2} \left[  \frac{1}{2}(t^2+\frac{1}{r^2})\p_t - t r \p_r -\frac{k}{r} \p_\phi \right],\\
Q_0 &=&  \p_\phi .\label{c00}
\eea
We have $\hat Q_0 = d\phi + k r dt$. The highest-weight scalars are given by 
\bea\label{phi00}
\Phi &=& \frac{1}{r},\qquad \lambda =e^{i \phi}.
\eea

\subsection{Global coordinates}

We denote global coordinates as $(\tau,y,\theta,\varphi)$. We have $\Psi = \varphi$.  The metric is 
\bea
ds^2 = \Gamma \left[  \frac{dy^2}{1+y^2} + d\theta^2 - (1+y^2)d\tau^2+\gamma^2 \left[ d\varphi + k yd\tau  \right]^2 \right]
\eea
We define
\bea
H_+ &=& i \frac{e^{i \tau}}{\sqrt{1+y^2}}(-y \p_\tau + i (1+y^2)\p_y -k \p_\varphi ) ,\\
H_0&=& i \p_\tau \\
H_- &=&i \frac{e^{-i \tau}}{\sqrt{1+y^2}}(-y \p_\tau - i (1+y^2)\p_y -k \p_\varphi ) ,\\
Q_0 &=& \p_\varphi .\label{c11}
\eea
We have $\hat Q_0 = d\varphi + k y d\tau $.  The highest-weight  scalars are
\bea\label{phi11}
\Phi &=& i \sqrt{2} \frac{e^{-i \tau} }{\sqrt{1+y^2}}, \qquad \lambda = e^{i \varphi +k \, \text{arctan}\, y} .
\eea

\subsection{Black hole coordinates}

We denote black hole coordinates as $(T,Y,\theta,\psi)$. We have $\Psi = \psi$.  The metric is 
\bea
ds^2 = \Gamma \left[  \frac{dY^2}{-1+Y^2} + d\theta^2 - (-1+Y^2)dT^2+\gamma^2 \left[ d\psi + k Y dT  \right]^2 \right]
\eea
We define
\bea
H_+ &=& \frac{e^{-T}}{\sqrt{-1+Y^2}}(Y \p_T +  (-1+Y^2)\p_Y -k \p_\psi ) ,\\
H_0&=&  \p_T \\
H_- &=& \frac{e^{T}}{\sqrt{-1+Y^2}}(Y \p_T -  (-1+Y^2)\p_Y -k \p_\psi ), \\
Q_0 &=& \p_\psi .\label{c22}
\eea
We have $\hat Q_0 = d\psi + k Y dT$. The highest-weight scalars are
\bea\label{phi22}
\Phi  &=& \sqrt{2} \frac{e^{T} }{\sqrt{-1+Y^2}},\qquad \lambda = e^{ i  \psi - i k  \, \text{arctanh}\, Y }.
\eea

\section{Canonical Euler potentials}
\label{cep}

The use of differential geometry in the covariant formulation of force-free electrodynamics was elegantly motivated and developed in \cite{Gralla:2014yja}. We will follow their conventions. Maxwell's equations are
\begin{equation} \label{Maxwell eq}
dF=0, \qquad d \star F= \star J
\end{equation}
where $F$ is the electromagnetic field 2-form, $J$ is the current 1-form and $\star$ is the Hodge star operator. The force-free condition is expressed as
\begin{equation} \label{FF cond}
J \wedge \star F=0.
\end{equation}
Force-free electromagnetic fields are degenerate, i.e., $\det(F)=0.$ This implies that the electromagnetic field can be written as
\begin{equation}
F=d\phi_{1}\wedge d\phi_{2}
\end{equation}
where $\phi_{1}$ and $\phi_{2}$ are the so-called Euler potentials which are determined up to a field redefinition of $(\phi_1,\phi_2)$ of unit Jacobian. 

We only consider spacetimes such that the metric admits a block diagonal form into a so-called two-dimensional Lorentzian toroidal part and a two-dimensional Euclidean poloidal part. The Kerr metric admits this decomposition and it  follows that the near-horizon extremal metric also does. 

Let us now concentrate on the near-horizon extremal Kerr in Poincar\'e coordinates. We anticipate that highest-weight solutions are constrained by $H_+ F = 0$ which is equivalent to requiring stationarity of the field strength. 
Canonical Euler potentials have been derived in this context for stationary and axisymmetric configurations  \cite{1997PhRvE..56.2198U} . Here, as a primer for describing the physical properties of highest-weight Poincar\'e solutions we find useful to first derive (complex) canonical Euler potentials for stationary but nonaxisymmetric configurations which are (complex) eigenstates of $\p_{\phi}$, $Q_0 F = i q F$.

The metric in Poincar\'e coordinates can be decomposed into the toroidal part spanned by $(t,\phi)$ with volume form $\eps_T = \Gamma \gamma r dt \wedge d\phi$, and the poloidal part spanned by $(r,\theta)$ with volume form $\eps^P = \frac{\Gamma}{r} dr \wedge d\theta$. We have $\eps = \eps^T \wedge \eps^P$, $\star \eps^T = -\eps^P$, $\star \eps^P = \eps^T$. 

\subsection{Stationary and axisymmetric case}

Let us first summarize the stationary and axisymmetric case as analyzed in \cite{1997PhRvE..56.2198U} and reviewed in \cite{Gralla:2014yja}. 
There is no toroidal electric field (and therefore no components of the field strength proportional to $dt \wedge d\phi$) for axisymmetric configurations as a simple consequence of Faraday's law.
We distinguish three scenarios:
\begin{description}
\item[generic case]  $i_{\p_\phi} F \neq 0$. One can choose
\bea\label{genpsi}
\phi_{1}=\psi(r, \theta), \qquad \phi_{2}=\phi+\psi_{2}(r, \theta) -\Omega(\psi) t.
\eea
The polar current $I(r,\theta)$ is defined as 
\bea
\star (d\psi \wedge d\psi_2 )= \frac{I(r,\theta)}{\sqrt{-g^T}}\eps^T.
\eea
It is equal to the electric current with respect to time $t$ flowing in the upward direction through the loop of revolution defined by the poloidal point $(r,\theta)$. Note that this interpretation breaks down beyond the velocity of light surface where $\p_t$ is spacelike. The force-free equations imply that $I = I(\psi(r,\theta))$. We therefore have
\begin{equation}
F=d\psi \wedge \bigl(d\phi - \Omega(\psi)dt\bigl)+I(\psi)\frac{dr\wedge d\theta }{\gamma r^{2}}.
\end{equation}
In particular, if $\Omega(\psi)=0$, there is no electric field, $i_{\p_t} F = 0$. 
\item[No poloidal magnetic field]  $i_{\p_\phi} F = 0 $, $i_{\p_t} F \neq 0$. One can choose instead
\bea
\phi_1  = \chi(r,\theta),\qquad \phi_2 = t+ \chi_2(r,\theta). 
\eea
We then define the polar current as $\star (d \chi \wedge d \chi_2) = \frac{I(r,\theta)}{\sqrt{-g^T}} \eps^T$ which has the same interpretation as above. The force-free equations imply $I=I(\chi(r,\theta))$. 
The corresponding field strength takes the form
\bea\label{Fs1}
F=d\chi \wedge dt + I(\chi)\frac{dr\wedge d\theta}{\gamma r^{2}}.
\eea

\item[Only toroidal magnetic field]  $i_{\p_t} F = 0$, $i_{\p_\phi} F = 0$. In that case,
\bea\label{scase}
\phi_1  = \chi(r,\theta),\qquad \phi_2 = \chi_2(r,\theta) ,\qquad F=I(\chi)\frac{dr\wedge d\theta}{\gamma r^{2}}.
\eea
There is no electric field and no poloidal magnetic field. 

\end{description}

\subsection{Stationary and $\p_\phi$-eigenvalue case}

Let us now consider a complex force-free field strength which is stationary, $\mathcal L _{\p_t}F=0$, and which is an $i q$ $\p_\phi$-eigenvalue, $\mathcal L_{\p_\phi} F = i q F$. 
Stationarity implies
\begin{align}
0 &= \mathcal{L}_{\p_t}F = d i_{\p_t} F= d\left(-\p_t \phi_2 d\phi_1 +\p_t \phi_1 d\phi_2\right)
\end{align}
where we used Cartan's formula, Bianchi's identity and the degeneracy of $F$.
By Poincar\'e's lemma, there exists a function $f=f(\phi_1, \phi_2)$ such that:
\bea \label{1symmy}
-\p_t \phi_2 d\phi_1 +\p_t \phi_1 d\phi_2=df. 
\eea
We distinguish here two cases (i) $i_{\p_t} F=df=0$ which implies that both Euler potentials are time independent and (ii) $i_{\p_t} F=df \neq 0$ to which we now turn our attention. 
Euler potentials are defined up to  the following arbitrariness: we may choose any other pair of potentials $(\tilde{\phi}_1, \tilde{\phi}_2)$, leaving the electromagnetic 2-form invariant, provided the map $(\phi_1, \phi_2) \rightarrow (\tilde{\phi}_1, \tilde{\phi}_2)$ has unit Jacobian determinant. Using this freedom, we choose $\tilde{\phi}_1=-f$. Let us check the existence of $\tilde{\phi}_2(\phi_1, \phi_2)$. The Jacobian of the transformation reads as
\begin{align}
1&=\frac{\p \tilde{\phi}_1}{\p \phi_1}\frac{\p \tilde{\phi}_2}{\p \phi_2}-\frac{\p \tilde{\phi}_1}{\p \phi_2}\frac{\p \tilde{\phi}_2}{\p \phi_1}=-\frac{\p f}{\p \phi_1}\frac{\p \tilde{\phi}_2}{\p \phi_2} + \frac{\p f}{\p \phi_2}\frac{\p \tilde{\phi}_2}{\p \phi_1}
\end{align}
which is a first order partial differential equation (PDE) for $\tilde{\phi}_{2}(\phi_1, \phi_2)$ and can be integrated with respect to $\phi_{2}$ if $\frac{\p f}{\p \phi_1}\neq 0$ or with respect to $\phi_{1}$ if $\frac{\p f}{\p \phi_2}\neq 0$.
With this new pair of Euler potentials, Eq. \eqref{1symmy} becomes
\bea
-\p_t \tilde{\phi}_2 d\tilde{\phi}_1 +\p_t \tilde{\phi}_1 d\tilde{\phi}_2=-d\tilde{\phi}_{1}
\eea
from which we read off the conditions
\bea
\p_t \tilde{\phi}_1=0, \qquad \p_t \tilde{\phi}_2=1
\eea
whose solutions are
\begin{equation}
\tilde{\phi}_1=\chi_1(r, \theta, \phi), \qquad \tilde{\phi}_2=t+\chi_2(r, \theta, \phi).
\end{equation}
Finally, merging cases (i) and (ii) and dropping tildes, Euler potentials for stationary solutions can be fixed to 
\begin{equation}
{\phi}_1=\chi_1(r, \theta, \phi), \qquad {\phi}_2=\eps\, t+\chi_2(r, \theta, \phi)
\end{equation}
where $\eps = 1$ if $i_{\p_t} F \neq 0 $ and $\eps = 0$ if $i_{\p_t} F = 0$. 

Let us now turn our attention to the second condition $\mathcal{L}_{\p_{\phi}}F=iqF$. We have
\begin{align}
0&=d i_{\p_{\phi}} F- iqF\nonumber \\
&=d i_{\p_{\phi}} (d\phi_1\wedge d\phi_2)-iq d\phi_1\wedge d\phi_2\nonumber\\
&=d \left[(i_{\p_{\phi}} d\phi_1)d\phi_2-(i_{\p_{\phi}} d\phi_2)d\phi_1\right]-iq d\phi_1\wedge d\phi_2 \label{eq65ff}\\
&=d\left[(\p_{\phi}\phi_1-iq\phi_1)d\phi_2-(\p_{\phi}\phi_2)d\phi_1\right]\nonumber\\
&=d\left[(\p_{\phi}\chi_1-iq\chi_1)(\eps dt+d\chi_2)-(\p_{\phi}\chi_2)d\chi_1\right]\nonumber
\end{align}
where we used Bianchi identity in the first step and stationarity in the last one. Let us first discuss the case $\eps  = 1$. 
Since $\chi_{1},\chi_2$ have no time dependence, from the identity
\bea
0=d(\p_{\phi}\chi_1-iq\chi_1) \wedge (dt+d\chi_2) - d(\p_{\phi}\chi_2)\wedge d\chi_1
\eea
we infer that 
\bea
\p_{\phi}\chi_1-iq\chi_1 =\mbox{const}, \qquad \p_{\phi}\chi_2=\kappa(\chi_{1})
\eea
where $\kappa(\chi_{1})$ is an arbitrary function of the Euler potential $\chi_{1}$ and where the arbitrary constant can be set to zero by shifting $\chi_1$.\\
From the first differential equation we have
\begin{equation}
\chi_1(r, \theta, \phi)= e^{iq\phi}\tilde{\chi}_1(r,\theta).
\end{equation}
From the second differential equation, we infer
\begin{equation}
\chi_2(r, \theta, \phi)=\int^{\phi}\kappa(e^{iq\phi'}\tilde{\chi}_1 )d\phi' +\tilde{\chi}_2(r, \theta).
\end{equation}
In conclusion, dropping the tildes for simplicity, the Euler potentials in the case $i_{\p_t} F \neq 0$ can be taken as
\begin{equation}
\boxed{
\phi_1=e^{iq\phi}\chi_1(r,\theta), \qquad \phi_2= t +\chi_2(r, \theta)+\int^{\phi}\kappa(e^{iq \phi'} \chi_1(r,\theta) )d\phi'
}\label{EP1}
\end{equation}
Let us compute the field strength. We define $h(r, \theta, \phi)=\int^{\phi}\kappa(e^{iq \phi'} \chi_1(r,\theta))d\phi'$, then
\bea
dh(r, \theta, \phi)=\frac{\p h}{\p r}dr + \frac{\p h}{\p \theta}d\theta + \frac{\p h}{\p \phi}d\phi
\eea
where 
\bea
\frac{\p h}{\p r}&=&\int^{\phi}\frac{\p \kappa(\phi_{1})}{\p \phi_{1}}|_{\phi \rightarrow \phi'}e^{iq\phi'}\p_{r}\chi_{1}d\phi', \quad \frac{\p h}{\p \theta}=\int^{\phi}\frac{\p \kappa(\phi_{1})}{\p \phi_{1}}|_{\phi \rightarrow \phi'} e^{iq\phi'}\p_{\theta}\chi_{1}d\phi', \\
\frac{\p h}{\p \phi}&=&\kappa(\phi_{1})
\eea
The exterior derivative of $\phi_{2}$ is 
\bea
d\phi_{2} =dt + d\chi_{2} + \kappa(\phi_{1})d\phi +\left(\int^{\phi}\frac{\p \kappa(\phi_{1})}{\p \phi_{1}}|_{\phi \rightarrow \phi'} e^{iq\phi'}d\phi'\right)d\chi_{1}
\eea
and the field strength takes the following form
\begin{equation}
F=d\phi_1 \wedge (dt+d\chi_2 ) + e^{iq\phi}\Biggl[\kappa(\phi_1)-iq\chi_1\int^{\phi}\frac{\p \kappa(\phi_{1})}{\p \phi_{1}}|_{\phi \rightarrow \phi'} e^{iq\phi'}d\phi'\Biggr]d\chi_1 \wedge d\phi.
\end{equation}

Let us now return to the case $\eps = 0$ ($i_{\p_t} F = 0$). We restart from \eqref{eq65ff}. By Poincar\'e's lemma there exists a function $f$ such that
\begin{equation}
\left[(\p_\phi \chi_1-iq\chi_1)(d\chi_2)-(\p_\phi \chi_2)d\chi_1\right]=df.
\end{equation}
If $df=0$ we find directly 
\bea
\p_\phi \chi_2 = 0,\qquad \p_\phi \chi_1-iq\chi_1 = 0
\eea
and we find the Euler potentials in the case $\p_t F = 0$ with $df=0$, 
\bea
\boxed{\phi_1=e^{iq\phi}\chi_1(r,\theta), \qquad \phi_2= \chi_2(r,\theta).}
\eea
In that case, the field strength is 
\begin{equation} \label{EP42}
F=e^{iq\phi}\Bigl(d\chi_1 \wedge d\chi_2  -iq\chi_1 d\chi_{2} \wedge d\phi\Bigr).
\end{equation}
If $df \neq 0 $, one can use again the ambiguity in the definition of Euler potentials to choose $\chi_1 = -f$. Then
\bea
\p_\phi \chi_2 = 1,\qquad \p_\phi \chi_1-iq\chi_1 = 0
\eea
and we find the Euler potentials $\phi_1=e^{iq\phi}\chi_1(r,\theta)$, $\phi_2= \phi +\chi_2(r,\theta)$. Since these potentials generalize \eqref{genpsi} when $\Omega =0$, we find it convenient to align the notations
so that finally we get in the case $\p_t F = 0$ with $df \neq 0$, 
\bea\label{case2}
\boxed{\phi_1=e^{iq\phi}\psi(r,\theta), \qquad \phi_2= \phi +\psi_2(r,\theta).}
\eea
In that case, the field strength is 
\begin{equation} \label{EP4}
F=e^{iq\phi}\Bigl(d\psi \wedge d\psi_2 + (d\psi -iq\psi d\psi_{2})\wedge d\phi\Bigr).
\end{equation}

\section{Maximally symmetric solutions}

Before considering highest-weight solutions, let us first obtain all $SL(2,\mathbb R) \times U(1)$ solutions to force-free electrodynamics. This analysis completes the one of \cite{Lupsasca:2014pfa}.

In the gauge $A_\theta = 0$, symmetries imply that the gauge potential has the form
\bea
A = A_0(\theta) \hat Q_0. 
\eea
This form is expressed only in terms of the metric and the $U(1)$ Killing vector and is therefore $SL(2,\mathbb R) \times U(1)$ invariant. Reality of the gauge potential requires that $A_0(\theta)$ be real. The current is given by
\bea
J=-\frac{\gamma}{\Gamma}\left[ \p_{\theta}\left(\frac{\p_{\theta}A_{0}}{\gamma} \right) + k^2\gamma A_{0} \right] \hat Q_0. 
\eea
The force-free condition is 
\bea\label{ff1}
0=J \wedge \star F = - \frac{\p_{\theta}A_{0}}{\Gamma}\left[ \p_{\theta}\left(\frac{\p_{\theta}A_{0}}{\gamma} \right) + k^2\gamma A_{0} \right]dt \wedge dr \wedge d\theta
\eea
Therefore the only force-free solution with nontrivial current is $A_0 \equiv -E_0/k$ constant which leads to 
\bea
A = - \frac{E_0}{k} \hat Q_0,\qquad F = -\frac{E_0}{k} d \hat Q_0,\qquad J = E_0 \frac{k\gamma^2}{\Gamma}\hat Q_0.\label{solSL2R}
\eea
The solution obeys
\bea
\star (F \wedge \star F) = -\frac{1}{2}F_{\mu\nu}F^{\mu\nu} = \frac{E_0^2}{\Gamma^2}
\eea
In the physical region $1-k^2\gamma^2 >0$ where the Killing vector $H_+=\sqrt{2} \p_t$ is timelike, the potential is electric with respect to the Killing time $t$. It is therefore electrically dominated and the plasma which underlies the current therefore moves at ultrarelativistic speed; see e.g. \cite{Komissarov:2004ms,Spitkovsky:2006np}. We will therefore not consider this solution further.  Note that these conclusions did not depend upon the Kerr functions \eqref{paramKerr}. It is a consequence of symmetry alone. 

\begin{figure}[htb] 
\centering
\includegraphics[width=12cm, height =6cm]{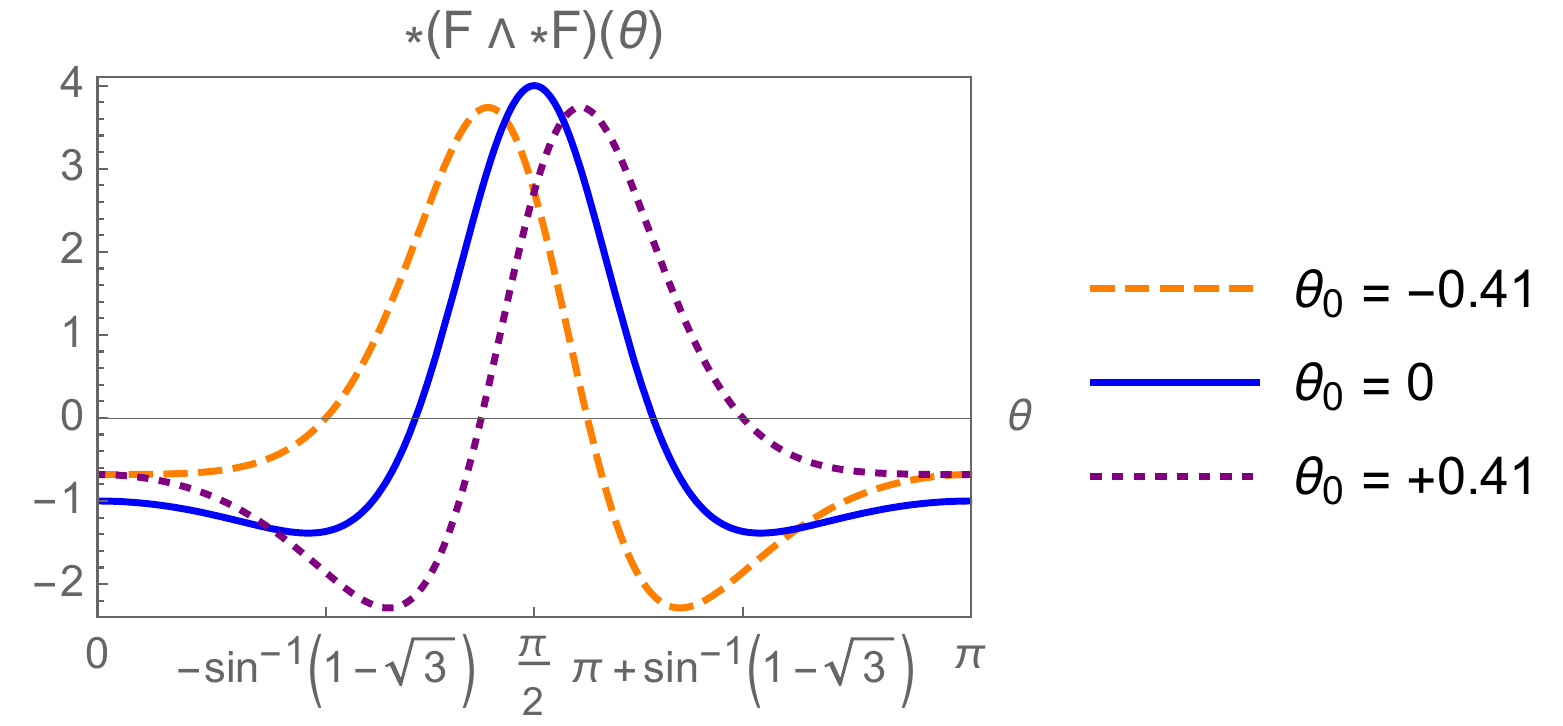}
\caption{$\star(F\wedge \star F) $ is negative outside the velocity of light surface, whose boundaries are highlighted, if the phase $\theta_{0}$ is in the range \eqref{range}.}
\label{fig: maxsym}
\end{figure}

The other solution to \eqref{ff1} has no current $J$ and therefore obeys free Maxwell's equations. It is given by
\bea
A &=& A_0(\theta) \hat Q_0 = M_0 \cos[\theta_0 -k \int^\theta d\theta' \gamma(\theta') ] \hat Q_0,\nonumber \\
&=& M_0 \cos[\theta_0 +2 \arctan \cos\theta] \hat Q_0,
\eea
where $\theta_0$ is a phase and $M_0$ is a constant magnitude. In the last line we specialized to the Kerr case \eqref{paramKerr}. The field strength is regular at the north and south poles since $A_0'(0)=A_0'(\pi)=0$.
The electromagnetic invariant is given by
\bea
\star(F\wedge \star F)= - \frac{1}{2}F_{\mu \nu}F^{\mu \nu} = \frac{1}{\gamma^2 \Gamma^2}\Bigl(-(\p_\theta A_0)^{2}+k^2 \gamma^2 A_0^2\Bigr).
\eea
We require that the field strength be magnetically dominated [$\star(F\wedge \star F) <0$] in the physical region, outside the velocity of light surface. It turns out that it is possible to do so upon choosing the phase in the range
\bea \label{range}
 -2 \arctan[\sqrt{2\sqrt{3}-3}]+\frac{\pi}{4} \leq \theta_0 \leq 2 \arctan[\sqrt{2\sqrt{3}-3}]-\frac{\pi}{4}
\eea
as clear from Fig. \ref{fig: maxsym}.

It is interesting that a maximally symmetric magnetic-type solution exists in the near-horizon region. However, it is not sustained by matter fields so it is unclear how such a vacuum solution would be compatible after backreaction with ``no hair theorems'' for stationary axisymmetric black hole solutions to Einstein-Maxwell theory.

\section{Highest-weight solutions}
\label{hwsol}

Our objective is to classify the (complex) highest-weight solutions of force-free electrodynamics in the near-horizon region of extremal Kerr, thereby completing the analysis of \cite{Lupsasca:2014hua}.
The first step in this program is to choose a basis for $SL(2,\mathbb R) \times U(1)$. There is a large automorphism group of this algebra which allows us to change basis while preserving the commutation relations \eqref{algebra}. Continuous automorphisms are parametrized by complex $(\alpha,\beta,\gamma,\delta)$ and given by
\bea
H_+ & \rightarrow &e^{-\gamma} \left[ (1+\alpha \beta)^2 H_+ +2\beta (1+\alpha \beta)H_0 +\beta^2 H_- \right],\\
H_0 & \rightarrow & \alpha (1+\alpha \beta)H_+ + (1+2\alpha \beta)H_0 + \beta H_-, \\
H_- & \rightarrow & e^\gamma \left[ \alpha^2 H_+ + 2\alpha H_0 + H_- \right],\\ 
Q_0 & \rightarrow & \delta \, Q_0.\label{auto}
\eea
The rescaling of $Q_0$ parametrized by $\delta$ can be absorbed into a rescaling of the angle $\phi$ and we do not consider it any further. Note that when $\alpha=-\beta^{-1}$, $e^\gamma = \beta^2$, $\delta =1$ and in the limit $\beta \rightarrow 0$ one finds the discrete automorphism given by
\bea
H_\pm \rightarrow H_\mp,\qquad H_0 \rightarrow -H_0 ,\qquad Q_0 \rightarrow Q_0. \label{lh}
\eea
Given a $SL(2,\mathbb R) \times U(1)$ basis, we look for solutions to the force-free equations \eqref{Maxwell eq}-\eqref{FF cond} satisfying the highest-weight conditions
\begin{equation} \label{HW def}
\begin{cases}
\mathcal{L}_{H_{+}}F=0\\
\mathcal{L}_{H_{0}}F=h F\\
\mathcal{L}_{Q_{0}}F=i q F
\end{cases}
\end{equation}
where $h$ is real or complex and $q$ is an integer. The choice of automorphism labeled by $\gamma$ is irrelevant: the resulting solutions to \eqref{HW def} will be identical. Also, it has been shown that the solutions for a given set $H_\pm,\, H_0$ and for a set transformed with real $\alpha,\beta$ can be mapped into each other with a real change of coordinates, which is an isometry of the metric. Therefore, one might just choose a particular basis and ignore such maps. 
However, the automorphisms labeled by complex $\alpha,\beta$ map a given solution to another nontrivial solution since they cannot be mapped into each other via a change of coordinates, as noted in \cite{Lupsasca:2014hua}. 
In summary, the automorphism group $SL(2,\mathbb C) = SL(2,\mathbb R) \times SL(2,\mathbb R)$ gets quotiented out by the real transformations $SL(2,\mathbb R)$ and the remaining real scaling transformations $\gamma \in U(1)$ and the resulting nontrivial choice of $SL(2,\mathbb R)$ basis is labeled by an element $(\alpha,\beta)$ of $SL(2,\mathbb R)/U(1)$. 

Since the force-free equations are nonlinear, several classes of solutions will appear and for each class, one could consider an extension of these solutions using the $SL(2,\mathbb R)/U(1)$ complex automorphisms. Let us now fix a $SL(2,\mathbb R)\times U(1)$ basis and classify the solutions. 

\subsection{$SL(2,\mathbb R)$ covariant basis of forms}

Since we would like to keep $SL(2,\mathbb R)$ covariance manifest, we need a $SL(2,\mathbb R)$ covariant definition of a basis of spacetime 1- and 2-forms that we will use to express all physical quantities. 
Also, since we describe highest-weight representations, it is computationally advantageous even though it is not necessary to choose a highest-weight basis. 

A basis of 1-forms $\mu^{i}$, $i=1,\dots 4$ of highest weight and weight 0, $\mathcal{L}_{H_{+}} \mu = \mathcal{L}_{H_{0}} \mu = 0$, is the following
\begin{align}
\mu^{1} &= k \gamma^2(\theta) \hat Q_0 - \frac{1}{\sqrt{2}\Gamma(\theta)} \Phi H_+,\\
\mu^{2} &= (1-k^2 \gamma^2(\theta)) \hat Q_0 + \frac{k}{\sqrt{2} \Gamma(\theta)} \Phi H_+, \\
\mu^{3} &= \frac{d\theta }{\gamma(\theta)},\\
\mu^{4} &= \frac{d\Phi}{\Phi},\label{basis1}
\end{align}
where $\Phi$ is the highest-weight scalar of weight 1 defined in \eqref{scalHW}. In Poincar\'e, global and black hole coordinates we have
\bea
\mu^{1} &=& r dt =y d\tau + \frac{i dy}{1+y^2} = Y dT + \frac{dY}{1-Y^2},\\
\mu^{2} &=& d\phi = d\varphi -k\frac{i dy}{1+y^2} = d\psi - k\frac{dY}{1-Y^2},\\
\mu^{3} &=& \frac{ d\theta }{\gamma(\theta)},\\
\mu^{4} &=& -\frac{dr}{r} = - i d\tau - \frac{ydy}{1+y^2}=dT + \frac{YdY}{1-Y^2}.
\eea
A basis of 2-forms $w^{i}$, $i=1,\dots 6$ of highest weight and weight 1, $\mathcal{L}_{H_{+}} w^{i}= \mathcal{L}_{H_{0}} w^{i} - w^{i} = 0$,  is given by 
\begin{align}
w^{1} &= \Phi \mu^{4}\wedge \mu^{3},\\
w^{2} &= \Phi \mu^{1}\wedge \mu^{3},\\
w^{3} &= \Phi k\gamma^2 {\hat{Q}_{0}}\wedge \mu^{3},\\
w^{4} &= \Phi d{\hat{Q}_{0}},\\
w^{5} &= \Phi {\hat{Q}_{0}}\wedge \mu^{4},\\
w^{6} &= \Phi \mu^{2}\wedge \mu^{1} = \frac{\Phi^2}{\sqrt{2}\Gamma} H_+ \wedge \hat Q_0.
\end{align}
This choice of basis is motivated by the Hodge duality properties
\begin{align}
\star w^{1} = -w^{6},\qquad 
\star w^{2} = -w^{5},\label{dual2}\qquad 
\star w^{3}= -w^{4},
\end{align}
and by the properties under the action of the total derivative 
\begin{equation}
dw^{i} = 0,\qquad \forall i \neq 3,\qquad
\star dw^{3}= \frac{k \Phi^2}{\sqrt{2}\Gamma^2} H_+.
\end{equation}
We could trade the basis elements $w^{3},w^{4}$ for another pair with $\star w^{3}= -w^{4}$ and $dw^{3}=0=d w^{4}$ but it would involve double integrals of $\gamma$ so we prefer not to use such a basis. The normalization of $\Phi$ in \eqref{phi11}-\eqref{phi22} was chosen so that the duality transformations \eqref{dual2} are obeyed.

Since the basis of 1- and 2-forms is real in Poincar\'e and black hole coordinates, these bases are well suited for studying real solutions as we will describe later on in Sec. \ref{physSol}. 

\subsection{Vector potential, field strength and current}

We expand the electromagnetic field tensor in the basis $w^{i}$
\begin{equation}
F_{(h, q)}(x)=F^{(h, q)}_i w^{i}
\end{equation}
where we chose
\bea
\mathcal L_{H_+} w^{i} = 0 ,\qquad \mathcal L_{H_0} w^{i} = w^{i}. 
\eea
After some simple algebra, the system \eqref{HW def} becomes
\begin{equation} \label{HW}
\begin{cases}
H_{+}F_{i}=0,\\
H_{0}F_{i}=(h-1) F_{i},\\
Q_{0}F_{i}=i q F_{i},
\end{cases}
\end{equation}
where $H_+ F_{i} \equiv  H_+^\mu \p_\mu F_{i}$, etc. 

We therefore reduced the problem to six decoupled scalar equations of the form \eqref{scalHW} whose solutions were provided in \eqref{solHW}. The most general expression for the electromagnetic field $F$ is therefore given by
\begin{equation} \label{F field}
F_{(h, q)}=\Phi_{(h-1, q)} f_{i}(\theta)w^{i}
\end{equation}
where $\Phi_{(h-1, q)}$ is the highest-weight scalar of weight $h-1$ and charge $q$. The Bianchi identities $dF=0$ reduce the number of independent functions $f_i(\theta)$ from six to three.
Then we must fix the remaining three functions by using the nonlinear equations of motion.

The most general vector potential $A_{(h,q)}$ generating $F_{(h,q)}$ will also contain three functions of $\theta$. We consider the following vector potential
\begin{equation} \label{potential}
A_{(h, q)}=\Phi_{(h, q)} a_{i}(\theta)\mu^{i}.
\end{equation}
Later on we will work in the gauge $a_4(\theta) = 0$ but we keep this function arbitrary in this section. 
The field strength functions $f_i(\theta)$ are then determined by
\begin{align}
\begin{cases}
f_{1}(\theta)&=ha_3(\theta)-\gamma a'_4(\theta),\\
f_{2}(\theta)&=-ikqa_3(\theta)+\gamma \left(-a'_1(\theta)+ka'_2(\theta)\right), \\
k\gamma^2 f_{3}(\theta)&=iqa_3(\theta)-\gamma a'_2(\theta),\\
kf_4(\theta)&=(1-h)a_1(\theta)+k\left(ha_2(\theta)-iqa_4(\theta)\right),\\
f_5(\theta)&=-ha_2(\theta)+iqa_4(\theta),\\
f_6(\theta)&=iqa_1(\theta).
\end{cases}\label{deff}
\end{align}
Of course, the Bianchi identities are then satisfied. Also, given an arbitrary set of field strength functions which obey the Bianchi identity, one can invert the system in a given gauge and solve for the gauge potential  functions. Any highest-weight field strength can then be expressed from a highest-weight gauge potential. 

From the field strength, one can obtain the current as
\begin{equation}
J_{(h, q)}=\Phi_{(h, q)} j_{i}(\theta)\mu^{i}
\end{equation}
where
\begin{align}
\begin{cases}
\gamma^2 \Gamma j_{1}&=\gamma f'_2+i q f_6 \left(k^2 \gamma^2-1\right)+k \gamma^2\left[k \gamma f'_3+\left(h-k^2 \gamma^2\right)f_4+(h-1) f_5\right ],\\
\Gamma j_{2}&=(h-1)f_5+i kqf_6-k\gamma\left( k \gamma f_4 -f'_3\right),\\
\Gamma j_{3}&=-(h-1)f_1-i k q (f_2+f_3),\\
\gamma^2 \Gamma j_4&=\gamma f'_1 - i q \left(k^2 \gamma^2 f_4+f_5\right).
\end{cases}\label{defj}
\end{align}

\subsection{Energy and angular momentum flux and variational principle}
\label{sec:E}

Let us pause to compute the energy and angular momentum flux. In Poincar\'e coordinates we define the energy with respect to $\p_t$, the angular momentum with respect to $-\p_\phi$ and the boundary is at $r \rightarrow \infty$ and the horizon is at $r = 0$. The flux density of energy and angular momentum per unit time per solid angle is given by
\bea
\dot{\mathcal E} & \equiv & \, \sqrt{-\gamma} T^{\mu}_{\;\; \nu} (\p_t)^\nu n_\mu = E(\theta) \, r^{2-2 h} e^{2 i q \phi} \label{EE}\\
\dot{\mathcal  J} &\equiv & \, -\sqrt{-\gamma} T^{\mu}_{\;\; \nu} (\p_\phi)^\nu n_\mu = J(\theta) \, r^{1-2h}e^{2 i q \phi}\label{JJ}
\eea
where $n^\mu$ is the unit normal, $\gamma_{\mu\nu}$ is the induced metric on constant $r$ surfaces and $T^{\mu\nu}$ is the electromagnetic stress tensor. For simplicity we only discuss complex fields. Real fields will be discussed in Section \ref{reality}. The functions $E(\theta) ,J(\theta) $ are explicitly given by
\bea
E(\theta) &=& h a_3(\theta) a_1'(\theta) + i q a_1(\theta) \left( (h-1) k \gamma a_1(\theta) + h \frac{1-k^2 \gamma^2}{\gamma} a_2(\theta) \right) ,\\
J(\theta) &=& h a_3(\theta) \left(-a_2'(\theta) +i \frac{q a_3(\theta)}{\gamma}\right)-i q a_1(\theta) \gamma \Big( (h-1) a_1(\theta) - h k a_2(\theta) \Big). \label{JJdef}
\eea
Requiring no energy and angular momentum flux at the boundary of the near-horizon region is, assuming $E(\theta) \neq 0$, equivalent to requiring $\text{Re} \,h > 1. $\footnote{This bound can also be obtained in global and black hole coordinates since these coordinate systems admit the same falloff at the spatial boundary.} The energy flux however then diverges at the Poincar\'e horizon.

Instead, let us define the energy with respect to the physical time from the point of view of the asymptotically flat observer. Let us remember how the near-horizon limit is taken. The outside asymptotically flat time,  radius and polar angle scale as $t_{out} = t/\lambda$, $r_{out} = M + \lambda r $, $\phi_{out} = \phi + \Omega_{ext} t/\lambda$ where $\lambda \rightarrow 0$. Here $M$ is the extreme Kerr mass and $\Omega_{ext}=\frac{1}{2M}$ is the extremal angular velocity. Then $\p_t^{out}+ \Omega_{ext} \p_\phi^{out} = \lambda \p_t$. Since the physical energy $\mathcal E_{out}$ and angular momentum $\mathcal J_{out}$ are associated with $\p_t^{out}$ and $-\p_\phi^{out}$, the physical deviation of energy flux density with respect to comoving flux density per asymptotically flat  time unit and solid angle can be expressed as
\bea \label{EEout}
\dot{\mathcal E}_{out} -\Omega_{ext}\dot{\mathcal J}_{out}  & = & \sqrt{-\gamma^{(out)}} T^{\mu(out)}_{\;\; \nu} (\p_t^{out}+ \Omega_{ext} \p_\phi^{out} )^\nu n^{(out)}_\mu 
\eea
where the right-hand side is a function of $(t_{out},r_{out},\phi_{out})$. We require that the stress-tensor admits a well-defined near-horizon limit. In order to reach stationarity in the near-horizon limit, we also impose that the stress-tensor only depends upon $(r_{out},\phi_{out} - \Omega_{ext} t_{out})$ in the near-horizon limit, i.e. the field co-rotates with the black hole at the horizon. Now when substituting $r_{out}$ as a function of $r$, all $r$ factors come equipped with a $\lambda$ factor as well. From the near-horizon scaling in Eq. \eqref{EE} we can deduce the scaling
\bea
\dot{\mathcal E}_{out}  -\Omega_{ext}\dot{\mathcal J}_{out}  \simeq \lambda^{2-2h} \dot{\mathcal E} \label{Eo}
\eea
Following the same reasoning, the angular momentum flux density scales as 
\bea
\dot {\mathcal J}_{out} \simeq \lambda^{1-2h}  \dot{ \mathcal J},\label{Jo}
\eea
The scaling limits \eqref{Eo}-\eqref{Jo} agree with \cite{Gralla:2016jfc}. 

Physical solutions should have a finite (noninfinite) physical energy and angular momentum flux in the limit $\lambda \rightarrow 0$. This requires $\text{Re}(h) \leq 1$ or $E(\theta) = 0$, $\forall \theta$. This condition also follows from requiring finiteness of the energy flux density \eqref{EE} at the Poincar\'e horizon in the near-horizon region. 
Angular momentum should also be finite in the asymptotically flat region. Comparing Eq. \eqref{Jo} with Eq. \eqref{JJ}, this is equivalent to requiring finiteness at the Poincar\'e horizon in the near-horizon region. This imposes that $\text{Re}(h) \leq \frac{1}{2}$ or $J(\theta) = 0$, $\forall \theta$.

Let us discuss the different cases similarly to \cite{Gralla:2016jfc}: 
\begin{itemize}
\item[(a)] $\text{Re}(h) < \frac{1}{2}$, one has $\dot{\mathcal E}_{out} =\dot{\mathcal J}_{out}=0$ and there is no energy flux extraction from the near-horizon region; 
\item[(b)] $\text{Re}(h) =\frac{1}{2}$, one finds $\dot{\mathcal E}_{out} = \Omega_{ext}\dot{\mathcal J}_{out}$ with  $ \dot{\mathcal J}_{out} \simeq  \dot{\mathcal J}$; 
\item[(c)] $\frac{1}{2} < \text{Re}(h) < 1$ and $J(\theta)=0$, one has $\dot{\mathcal E}_{out} =\Omega_{ext}\dot{\mathcal J}_{out}$ with $\dot{\mathcal J}_{out}$ (and hence $\dot{\mathcal E}_{out}$) undetermined;
\item[(d)] $\text{Re}(h) = 1$ and $J(\theta)=0$ one has $\dot{\mathcal E}_{out} -\Omega_{ext}\dot{\mathcal J}_{out} \simeq \dot{\mathcal E}$ with $\dot{\mathcal J}_{out}$ undetermined;
\item[(e)] $\text{Re}(h) >1$ and $J(\theta)=E(\theta)=0$ and $\dot{\mathcal E}_{out}$ and $\dot{\mathcal J}_{out}$ are not determined from the near-horizon region information.
\end{itemize}
Here we allow for complex $h$, which indicates an oscillatory behavior at the spatial boundary. In the case (b) and (c) the energy flux saturates the superradiant bound. This indicates that the field exactly corotates with the black hole. It can only happen for photons orbiting at the Killing horizon which classically never escape it. This is a feature of the ideal extremal solution.  In the case (d) one has $\dot{\mathcal E} \leq 0$. Indeed, as clearly explained in \cite{Gralla:2014yja}, upon changing the mass and angular momentum of the black hole as $\delta M = -\delta \mathcal E_{out}  $ and $\delta J = - \delta \mathcal J_{out}$ with outgoing radiation we must obey $\delta M - \Omega_H \delta J = \frac{\kappa}{8\pi}\delta A \geq 0$. Therefore $\dot{\mathcal E}_{out}  \leq  \Omega_{ext}\dot{\mathcal J}_{out} $. This physically corresponds to the fact that an outflow of energy necessarily comes accompanied with an outflow of angular momentum. This inequality can also be obtained from the null energy condition $T_{\mu\nu}\chi^\mu \chi^\nu \geq 0 $ using the Killing horizon generator $\chi = \p_t + \Omega_H \p_\phi$.  

We will be interested in solutions which admit an energy flux in the asymptotic region which implies $\text{Re}(h) \geq \frac{1}{2}$. We will limit ourselves to near-horizon solutions which bring information about the asymptotically flat region, which implies  $\text{Re}(h) \leq 1$.  Our definition of potentially physical solutions in Section \ref{listphys} will only refer to solutions with 
\bea
\boxed{\frac{1}{2} \leq \text{Re}(h)\leq 1 \qquad \text{and} \qquad \left(\text{Re}(h)-\frac{1}{2}\right)J(\theta) = 0,} \label{boundh}
\eea
which covers the cases (b), (c) and (d).

The bound \eqref{boundh} also has a special meaning in the variational principle. Indeed, the action of force-free electrodynamics is given by
\bea
S^{FF} = -\frac{1}{2}\int d\phi_1 \wedge d\phi_2 \wedge  \star (d\phi_1 \wedge d\phi_2)
\eea
whose variation gives a boundary term $\int_I \Theta^\mu (d^3x)_\mu$ with
\bea
\Theta^\mu = \sqrt{-g}F^{\mu\nu} (\delta \phi_2 \p_\nu \phi_1 - \delta \phi_1 \p_\nu \phi_2)
\eea
where $F_{\mu\nu} = \p_\mu \phi_1 \p_\nu \phi_2 - \p_\mu \phi_2 \p_\nu \phi_1 $ is the field strength. It is easy to check that for a highest-weight solution one has
\bea
\Theta^r \simeq r^{1-2h}.
\eea
Therefore the case $\text{Re} \,h = \frac{1}{2}$ is precisely the one for which there is symplectic flux, and additional care is required to define the variational principle. The variational principle is otherwise well-defined in the range \eqref{boundh}.

\subsection{Solving the force-free condition}
\label{sec:solv}

For the highest-weight ansatz, the force-free condition \eqref{FF cond} is equivalent to the following relations between $f_i$'s and $j_i$'s:
\begin{equation} \label{FF syst}
	\begin{cases}
	k(f_2+f_3)j_3+(f_5+k^2\gamma^2f_4)j_4=f_6j_1,\\
	f_2j_3+k\gamma^2f_4j_4=f_6j_2,\\
	-k\gamma^2f_4j_1 +(f_5+k^2\gamma^2f_4)j_2=f_1j_3,\\
	f_2j_1-k(f_2+f_3)j_2=f_1j_4.
	\end{cases}
\end{equation}
Now, we see that we can recast this system of equations in the following form
\begin{equation} \label{FF syst2}
\begin{bmatrix}
\mathbb{A} & -f_{6} 1\!\!1 \\
-f_{1}1\!\!1 & \mathbb{B}\\
\end{bmatrix}
\begin{bmatrix}
\bold{x}\\
\bold{y}\\
\end{bmatrix}
=
\begin{bmatrix}
0\\
0\\
\end{bmatrix}
\end{equation}
where
\begin{equation}
\mathbb{A}=
\begin{bmatrix}
k(f_{2}+f_{3}) & f_{5} +k^{2}\gamma^{2}f_{4} \\
f_{2} & k\gamma^{2}f_{4} \\
\end{bmatrix},\quad 
\bold{x}= 
\begin{bmatrix}
j_{3}\\
j_{4} \\
\end{bmatrix},\quad 
\bold{y}= 
\begin{bmatrix}
j_{1}\\
j_{2} \\
\end{bmatrix}
\end{equation}
and 
\bea
\mathbb{B}=\sigma \mathbb{A}^{T}\sigma, \qquad \sigma = \left( \begin{array}{cc} 0&1 \\-1 &0 \end{array} \right). 
\eea

The linear system \eqref{FF syst2} has nontrivial solutions if its determinant is vanishing, i.e.,
\begin{equation}
0=\det 
\begin{bmatrix}
\mathbb{A} & -f_{6} 1\!\!1 \\
-f_{1}1\!\!1 & \mathbb{B}\\
\end{bmatrix}
=[\det(\mathbb{A})+f_{1}f_{6}]^{2},
\end{equation}
which turns out to be equivalent to the degeneracy condition of the field strength $F$. For definiteness, we work in the gauge
\bea
a_4 = 0 \quad \forall h \qquad \text{and} \qquad a_3 =a_4 =  0 \quad \text{for $h=0$}.
\eea
The degeneracy condition can then be written as 
\bea \label{deg00}
(h-1)a_1 a_2'- h a_1' a_2 +iq \frac{a_1a_3}{\gamma} = 0. 
\eea
Substituting \eqref{deff} and \eqref{defj} in \eqref{FF syst} one obtains in total three nonlinear coupled ODEs in terms of the gauge potential functions $a_1,a_2,a_3$. The first equation is \eqref{deg00}. The two other equations are lengthy and unenlightening. Since an interested reader can easily reproduce them we will omit them here. 

Given the difficulty of these equations, it is useful to organize the solutions by first classifying the solutions in terms of field strength functions only. 
After analysis, we could branch these nonlinear equations into seven independent and complete (still nonlinear) subcases where they can be solved:
\begin{enumerate}
\item $f_1=f_2=f_3=f_4=f_6=0$, $f_5 \neq 0$, $j_2=j_4=0$,
\item $f_1=f_2=f_3=f_6=0$, $f_4 \neq 0 $, $j_4=0$, $j_1 = k j_2 + \frac{f_5 j_2}{k f_4 \gamma^2}$,
\item $f_1=f_2=f_4=f_6=0$, $f_3 \neq 0$, $j_2=0$, $j_3 = -\frac{f_5 j_4}{k f_3}$,
\item $f_1 = 0$, $f_2 \neq 0$, $f_5 = \frac{k^2 \gamma^2 f_3 f_4}{f_2}$, $j_1 = k j_2+ \frac{k f_3 j_2}{f_2}$, $j_3 = \frac{f_6 j_2 - k \gamma^2 f_4 j_4}{f_2}$,
\item $f_1 =f_2=f_3 = 0$, $f_6 \neq 0$, $j_1 = \frac{j_4(f_5+k^2 \gamma^2 f_4)}{f_6}$, $j_2 = \frac{k \gamma^2 f_4 j_4}{f_6}$,
\item $f_1 = f_2 = f_4=0$, $f_6 \neq 0$, $j_2=0$, $j_1 = \frac{k f_3 j_3+f_5 j_4}{f_6}$,
\item $f_1 \neq 0$, $f_6 = \frac{f_2 f_5 -k^2 \gamma^2 f_3 f_4 }{f_1}$, $j_3 = \frac{f_5 j_2 + k \gamma^2 f_4 (-j_1+k j_2)}{f_1}$, $j_4 = \frac{-k f_3 j_2 +f_2 (j_1 - k j_2)}{f_1}$.
\end{enumerate}
Substituting \eqref{deff} and \eqref{defj} one obtains the nonlinear equations in terms of the gauge potential functions $a_i$, $i=1,2,3$. We were able to fully solve the cases 1 to 6. However, the general solution in case 7 eluded us. The resulting two ODEs are strongly nonlinear and it is not clear whether we obtained all possible solutions. We will present in Sec. \ref{allsol} all solutions found.

\subsection{Electromagnetic types}

In order to better present the solutions, it is useful to first present some criteria for distinguishing magnetically dominated solutions from null solutions and electrically dominated ones. 

We define the (not normalized) electric and magnetic fields with respect to the vector $v^\mu$ as 
\bea 
E_\mu=  F_{\mu\nu}v^{\nu}, \qquad \quad \qquad B_\mu=(\star F)_{\mu\nu}v^{\nu},
\eea
where we require the vector $v^{\mu}$ to be timelike close to the north and south poles, in the physical region $-1+k^{2}\gamma^{2} < 0$, so that it can be tangent to an observer. If we restrict our attention to real fields,  $E_\mu$ and $B_\mu$ are two spatial vectors in the region $-1+k^{2}\gamma^{2} < 0$ since $E_\mu v^\mu=0=B_\mu v^\mu$.  They can however be timelike beyond the velocity of light surface.
The degeneracy condition on the electromagnetic field is given by
\begin{align}\label{deg0}
0&=\star(F \wedge F)=\frac{1}{4}\epsilon^{\mu\nu\gamma\delta}F_{\mu\nu}F_{\gamma\delta}=\frac{2}{v\cdot v} E_{\mu}B^{\mu},
\end{align}
where we used
$
F_{\mu\nu}= \frac{1}{v\cdot v}\left(2 E_{[\mu}v_{\nu]} - \epsilon_{\mu\nu\gamma\delta}B^{\gamma}v^{\delta} \right)
$. 
The second electromagnetic invariant 
\begin{align}
I_2&=\star(F \wedge \star F)=-\frac{1}{2} F_{\mu\nu}F^{\mu\nu}=-\frac{1}{v\cdot v}(E_{\mu}E^{\mu}-B_{\mu}B^{\mu})
\end{align}
tells us whether the solution is magnetically dominated $I_2 <0$, null $I_2=0$ or electrically dominated $I_2 > 0$. The criteria for having a drift velocity of charged particles less than the speed of light is to have magnetic dominance $I_2 < 0$; see e.g. \cite{Komissarov:2004ms}.

For the (complex) highest-weight ansatz, the degeneracy condition \eqref{deg0} is written as \eqref{deg00}. In order to get some intuition it is useful to first concentrate on Poincar\'e coordinates $(t,r,\theta,\phi)$. The field strength then reads as
\bea
F= \frac{e^{i q \phi}}{r^h} \left (
\begin{array}{cccc}
0 & (h-1)a_1(\theta) & -ra'_1(\theta) & -iqra_1(\theta) \\
  & 0 & -h \frac{a_3(\theta)}{\gamma r} & -h \frac{a_2(\theta)}{r} \\
  &  & 0 & a'_2(\theta)-iq \frac{a_3(\theta)}{\gamma}\\
  &  &    & 0 
\end{array}
\right ).
\eea
It is therefore natural to distinguish four (partially overlapping) qualitative classes of solutions types\footnote{Poincar\'e magnetic was denoted as Type M in \cite{Lupsasca:2014hua} while Poincar\'e electric was denoted as Type E and Poincar\'e generic was denoted as Type E-M. Our terminology emphasizes the role of the Poincar\'e time $t$ in the $3+1$ decomposition.}:
\begin{description}
\item{\bf Poincar\'e magnetic}  $ \Leftrightarrow a_1=0$ 
\item{\bf Poincar\'e electric} $ \Leftrightarrow a_2 = a_3=0$
\item{\bf Poincar\'e nontoroidal} $ \Leftrightarrow a_{2} = 0$, $q = 0$
\item{\bf Poincar\'e generic} $ \Leftrightarrow a_{1} \neq 0 $ and $a_{2} \neq 0$ which obey \eqref{deg00}. 
\end{description}

A Poincar\'e magnetic solution has no electric field with respect to $\frac{\p}{\p t}$. Any such real solution is therefore magnetically dominated. For example, an axisymmetric configuration ($q=0$) with real weight $h$ and $a_i$'s is real and magnetically dominated. (Other real solutions can be obtained by superpositions as will be discussed in Sec. \ref{reality}.) A Poincar\'e electric solution has no magnetic field with respect to the $3+1$ decomposition involving the Poincar\'e time $t$. Any such real solution is therefore electrically dominated. A Poincar\'e nontoroidal solution has no components of the electromagnetic field along $d \phi$. This implies that the electric field has no toroidal components while the magnetic field (related to the dual of $F$) has no poloidal components. Since the toroidal and poloidal subspaces are orthogonal, it is indeed consistent with $E_{\mu} B^{\mu} = 0$. In general, there are still toroidal magnetic fields and poloidal electric fields but nothing prevents us from canceling one such field. The solution can then also be either Poincar\'e magnetic or Poincar\'e electric. The Poincar\'e generic solution has no particular electromagnetic property with respect to Killing time $t$. For real fields, there might however be another observer that identifies the solution as magnetically or electrically dominated or null. In fact, we will encounter such cases in Sec. \ref{physSol}.

Let us mention that this classification extends to black hole coordinates. A Poincar\'e magnetic solution will also be magnetically dominated in the physical region in the black hole patch. Indeed, the vector field
\begin{align}
v&=\frac{\sqrt{\Gamma(\theta)}}{\sqrt{|-1+k^2\gamma(\theta)^2|}}\Bigl((k^2\gamma(\theta)^2-1)\mu^{1}+k\gamma(\theta)^2\mu^{2}\Bigr)
\end{align}
is real and timelike in the physical region ($v \cdot v =-\mbox{sign}(1-k^2\gamma(\theta)^2)$) and it lies in the kernel of $F$. Observers tangent to that vector field see therefore no electric field. 
Also, a Poincar\'e electric solution admits a real vector in the black hole patch which lies in the kernel of $\star F$ and which is timelike. The 1-form is given by
\bea
v=\sqrt{\Gamma(\theta)}\mu^{1}
\eea
whose norm is $-1$ everywhere.

\subsection{$SL(2,\mathbb R)$ descendants}

Even though the force-free electrodynamics equations are nonlinear, they admit under certain circumstances a linear superposition principle. Several sufficient conditions for linear superposition were thoroughly discussed in \cite{Lupsasca:2014pfa,Lupsasca:2014hua}. We review some key propositions here. First, two solutions $A_1$ and $A_2$ can be linearly superposed if their respective currents $J_1$, $J_2$ are collinear, $J_1 \sim J_2$. Second, a solution $A$ can be linearly superposed with its descendant $\mathcal L_K A$ associated with a Killing vector  $K$ if the descendant of the current is collinear with the current itself, $\mathcal L_K J \sim J$. Moreover, under the same assumption $\mathcal L_K J \sim J$,
the gauge field $P(\mathcal L_K) A$ is also a solution  for any polynomial $P$. While these conditions might not be necessary for linearly superposing solutions, they allow us to build large classes of solutions. 

In the following, we will check whether solutions with different weight $h$ and $U(1)$ charge $q$ can be superposed simply by looking at the current, which will be computed in Appendix \ref{ID}. If they obey the criteria that currents with different values of $(h,q)$ are collinear we will be able to linearly superpose them. 

$SL(2,\mathbb R)$ descendants of the highest-weight solution $A$ are defined from acting with $\mathcal L_{H_-}$ on $A$. These gauge fields are not necessarily solutions, except in special cases. From the above propositions, descendants of solutions having a current $J$ with components along a linear combination of $Q_0$, $H_{-}$ and $\mu^{3}=d\theta/\gamma(\theta)$ will be solutions because such vector fields commute with $H_-$.
 Another possibility to linearly build a superposition of descendants is to start from a highest-weight solution with $J \sim H_0$. Then since $[H_-,H_0] \sim H_0$, $\mathcal L_{H_-}J \simeq J$. We will qualify  solutions which obey the property $\mathcal L_{{H_{-}}} J \propto J$ as ``{\bf admitting descendants}.''

\subsection{List of solutions}
\label{allsol}

Here, we list all solutions to force-free electrodynamics with nonvanishing current that we found in our analysis starting from the highest-weight ansatz. We first classify the solutions according to their highest-weight representation labeled by the (complex) weight $h$ and the (integer) $U(1)$-charge $q$ and then by their Poincar\'e electromagnetic type.
The functions $X_i(\theta)$, $i=1,2,3,4,5$ obey ODEs in $\theta$ which are described in Appendix \ref{app:ODE}. More details on the solutions including the field strength, current and Euler potentials can be found in Appendix \ref{ID}.

\begin{description}
\item[(h, q)-eigenstates] Two classes of solutions with arbitrary weight $h$ and $U(1)$-charge $q$:

\begin{description}
\item[$\bullet$ Poincar\'e magnetic]
\begin{equation}\label{solX5}
A=\int dh \sum_{q \in \mathbb Z} \Phi^{h}\lambda^{q}\left[ X_{5}\mu^{2}-\frac{ i q\gamma(1-k^{2}\gamma^{2})}{q^{2}-\Delta(h,q)\gamma^{2}} X'_{5}\mu^{3} \right]
\end{equation}
where $X_5=X_5(\theta; h ,q)$ and $\Delta(h,q)=h(h-1)+k^2 q^2$. The solution is pure gauge for $h=0$. When $q=0$, $h \neq 0,1$ the solution reduces to \eqref{eq: 11}.

\item[$\bullet$ Poincar\'e generic]
\begin{equation}\label{X2a}
A_{(h, q)}=\Phi^{h}\lambda^{q}\Bigl[h(h-1)X_2 \mu^1-kq^2 X_2\mu^2+i kq\gamma X_2'\mu^3\Bigr]
\end{equation}
where $X_2=X_2(\theta;\Delta(h,q), c_1=q^2)$. The solution is pure gauge for $h=0$. 
\end{description}

\end{description}

\begin{description}
\item[($\bf h \neq 0, q=0$)-eigenstates] Four classes of axisymmetric solutions with arbitrary weight $h$ and one special subcase:

\begin{description}
\item[$\bullet$ Poincar\'e generic]
\begin{equation} \label{eq: 3}
A_{(h, 0)}=c_1^h \Phi^{h}\Bigl[- X_{3}^{h-1}\mu^{1}+ X_{3}^h\mu^{2}\pm \sqrt{\xi}  X_{3}^{h-1}\mu^{3}\Bigr]
\end{equation}
where $X_3 = X_3(\theta; h ,\xi)$, $h,\xi \in \mathbb C$ and $c_1 \neq 0$.

\item[$\bullet$ Poincar\'e magnetic] 
\begin{equation} \label{eq: 3b}
A_{(h, 0)}= c_2^h \Phi^{h}\Bigl[  X_{4}\mu^{2}\pm  X_{4}^{\frac{h-1}{h}}\mu^{3}\Bigr]
\end{equation}
where $X_4 = X_4(\theta;\Delta(h) )$ with $\Delta(h) = h(h-1)$ and $c_2 \neq 0$. 

\item[$\bullet$ Poincar\'e magnetic] 
\begin{equation} \label{eq: 11}
A=\int dh\,  \Phi^{h}   X_{1} \mu^{2} 
\end{equation}
where $X_1=X_1(\theta ; \Delta(h))$.

\item[$\bullet$ Poincar\'e nontoroidal]
\begin{equation}
A_{(h, 0)}=\Phi^{h} X_{2} \Bigl[h \mu^{1} \pm\sqrt{c_{1}}\mu^{3}\Bigr] \label{eq:12}
\end{equation}
where $X_2=X_2(\theta,\Delta(h,q),c_1)$. 

\item[$\bullet$ Poincar\'e electric and nontoroidal - admitting descendants]
\begin{equation} \label{eq: AD1}
A_{(h, 0)}=\Phi^{h} X_{2}\mu^{1} 
\end{equation}
where $X_2=X_2(\theta,\Delta(h,q),0)$. It is the special case $c_1 = 0$ of \eqref{eq:12}.
\end{description}

\end{description}

\begin{description}
\item[$\bf (h= 0,q \neq 0$)-eigenstates] One weight 0 solution with arbitrary $U(1)$ charge $q$ :

\begin{description}

\item[$\bullet$ Poincar\'e electric] 
\begin{equation} \label{eq: 13bis}
A_{(0, q)}=\lambda^{q} e^{\pm\int \frac{q}{\gamma}d\theta}\mu^{1}.
\end{equation}
\end{description}
\end{description}

\begin{description}
\item[$\bf (h= 1,q \neq 0$)-eigenstates] One weight $1$ solution with arbitrary $U(1)$ charge $q$:

\begin{description}
\item[$\bullet$ Poincar\'e electric - admitting descendants]
\begin{equation} \label{eq: 14}
A=\sum_{q \in \mathbb Z} \Phi \lambda^{q} e^{\pm\int \frac{q}{\gamma}d\theta}\mu^{1}.
\end{equation}
\end{description}
\end{description}

\begin{description}
\item[$\bf (h=\pm {i}kq,q \neq 0$)-eigenstates] Two weight $\pm {i}kq$ solutions with arbitrary $U(1)$ charge $q$:

\begin{description}
\item[$\bullet$ Poincar\'e generic]
\begin{align} \label{eq: 15}
A_{(h=\pm {i}kq, q)}&=\Phi^{h}\lambda^{q}e^{\pm\int\frac{d\theta}{\gamma}}\Bigl[{i}kq\mu^{1}+{i}q \mu^{2}\pm\mu^3\Bigr],\\ \label{eq: 16}
A_{(h=\pm {i}kq, q)}&=\Phi^{h}\lambda^{q}\Bigl[k\mu^{1}+\mu^{2}\Bigr].
\end{align}
\end{description}
\end{description}

\begin{description}
\item[$\bf (h= 1 \pm {i}kq,q \neq 0$)-eigenstates] One weight $1\pm {i}kq$ solution with arbitrary $U(1)$ charge $q$:

\begin{description}
\item[$\bullet$ Poincar\'e generic - null] 
\begin{equation}\label{eq: 17}
\sum_{q \in \mathbb Z} A_{(h(q)=1 \pm {i}kq, q)}=\sum_{q \in \mathbb Z}  \Phi^{h}\lambda^{q}\Bigl[h(q)a_1(\theta)\mu^{1}\pm {i}qa_1(\theta)	\mu^{2}\pm\gamma a_1'(\theta)\mu^3\Bigr].
\end{equation}
\end{description}
\end{description}

\begin{description}
\item[$\bf (h= 1, q = 0$)-eigenstates] Two weight $1$ axisymmetric solutions:

\begin{description}
\item[$\bullet$ Poincar\'e nontoroidal - null]
\bea \label{eq: 18}
A_{(1, 0)}=\Phi \Bigl[a_{1}(\theta)\mu^{1}\pm \sqrt{c_{3}+[\gamma a_{1}'(\theta)]^{2}}\mu^{3}\Bigr]
\eea
where $a_{1}$ is an arbitrary function.

\item[$\bullet$ Poincar\'e magnetic]
\bea \label{eq: 11b}
A_{(1,0)} = \Phi (c_2 \mu^2+c_3 \mu^3) .
\eea
\end{description}
\end{description}

\begin{description}
\item[$\bf (h= 0, q = 0$)-eigenstates] One weight $0$ axisymmetric solution:

\begin{description}
\item[$\bullet$ Poincar\'e electric and nontoroidal - admitting descendants]
\bea\label{eq: 19b}
A_{(0,0)} = -E_0 ( \mu^1+\frac{1}{k}\mu^2).
\eea
This solution is just the $SL(2,\mathbb R)$ invariant solution \eqref{solSL2R}. 
\end{description}

\end{description}

In comparison with \cite{Lupsasca:2014hua}, the solutions \eqref{eq: 3b}, \eqref{eq:12}, \eqref{eq: 15}, \eqref{eq: 16}, \eqref{eq: 17}, \eqref{eq: 18} are new and the solutions   \eqref{eq: 13bis}, \eqref{eq: 14} are given with two branches distinguished by a sign.

\section{Potentially physical near-horizon solutions}
\label{physSol}

So far, we obtained a list of complex solutions to the force-free equations with various types of electromagnetic fields. Physical solutions should obviously be real. In this section, we will first discuss how to build real solutions from complex solutions. 

Force-free fields are sustained by matter which should travel at subluminal speed or at the speed of light in the case of photons. It forces the field strength to be magnetically dominated or null. Since the region beyond the velocity of light surface is not physical, we will only require that physical solutions be magnetically dominated or null in the polar region between the poles and the velocity of light surface. We will refer to such real solutions as the near-horizon solutions which we will list  in Sec. \ref{sec:near}. Finally, solutions should have finite and computable energy and angular momentum flux with respect to the asymptotically flat observer and should be solutions to the variational principle. According to the analysis of Sec. \ref{sec:E}, it leads to the bound \eqref{boundh}. The latter set of solutions will be refered to as the potentially physical near-horizon solutions, described in Sec. \ref{listphys}.

\subsection{Reality conditions}
\label{reality}

In Poincar\'e and in black hole coordinates the basis $\mu^{i}$ is real and therefore real solutions are obtained for gauge potentials with real components along the basis functions. 
Since $q \in \mathbb Z$, $h \in \mathbb C$ we have 
\bea
A_{(h,q)}^* = \Phi_{(h^*,-q)} a^*_i(\theta) \mu^{i}
\eea
Real solutions can be obtained in various ways. 

First, if $h$ is real and $q=0$, then $a_i(\theta)$ is real and the solution is therefore real. 

Second, if the current $J$ and its complex conjugate $J^*$ are proportional to each other, then one can linearly superpose these solutions to obtain a solution and therefore the real and imaginary parts of the gauge field are also solutions to the force-free equations. 

Finally, if  the current $J$ and its complex conjugate $J^*$ are not proportional to each other, one might attempt to find a nonlinear superposition of the solution and its complex conjugate with additional nonlinear terms but there is no known systematic procedure to do so. 

\subsection{List of near-horizon solutions}
\label{sec:near}

Let us now list all real magnetically dominated or null solutions that we could build from the complex solutions enumerated in Sec. \ref{allsol}. At this stage, we list these solutions with arbitrary highest-weight $h$.

\begin{itemize}
\item Nonaxisymmetric, magnetic:
\begin{equation}\label{solX5real}
A^M =\int dh \sum_{q \in \mathbb Z} \Phi^{h}\lambda^{q}\left[ X_{5}\mu^{2}-\frac{ i q\gamma(1-k^{2}\gamma^{2})}{q^{2}-\Delta(h,q)\gamma^{2}} X'_{5}\mu^{3}\right] +c.c. 
\end{equation}
where $X_5=X_5(\theta; h ,q)$ and $\Delta(h,q)=h(h-1)+k^2 q^2$. The axisymmetric case $q = 0$ is listed below and we have then $X_5(\theta,h,0)=X_1(\theta, \Delta(h))$.

\item Nonaxisymmetric, magnetic:
\begin{eqnarray}
A^{EM}_{(h=1+i \mu , q)} &=& \Phi^{1+i \mu }\lambda^q\Bigl[h(h-1)X_2 \mu^1-kq^2 X_2\mu^2+i kq\gamma X_2'\mu^3\Bigr]\nn \\
&& +\Phi^{1-i \mu}\lambda^{* q} \Bigl[h^*(h^*-1)X^*_2 \mu^1-kq^2 X^*_2\mu^2-i kq\gamma X^{* \prime}_2\mu^3\Bigr]\label{eq56}
\end{eqnarray}
where $X_2=X_2(\theta;\Delta(h,q), c_1=q^2)$. The solution is magnetically dominated ($I_2 < 0$) in the range $- k q < \mu < k q$. The borderline case $\mu^2=k^2 q^2$ is a null solution. (This is an example of Poincar\'e generic solution which is magnetically dominated.)

\item Axisymmetric {(magnetic dominance not checked)}:
\begin{equation}
A^{EM}_{(h, 0)}=c_1^h \Phi^{h}\Bigl[- X_{3}^{h-1}\mu^{1}+X_{3}^h\mu^{2}\pm \sqrt{\xi}X_{3}^{h-1}\mu^{3}\Bigr]
\end{equation}
where $h > 1$, $c_1$ is real, and $\xi > 0$. (In the case $h=1$, $X_3$ becomes $X_1$ and the solution is not smooth at the velocity of light surface; see Appendix \ref{ID}). It has been observed that for $h=-1$ this solution is magnetically dominated \cite{Zhang:2014pla}. We did not check if it is the case for $h \geq 1$. The solution to the nonlinear equation for $X_3(\theta)$ is required which we did not obtain here.

\item Axisymmetric, magnetic:
\begin{equation} 
A^{M}_{(h, 0)}= c_2^h \Phi^{h}\Bigl[  X_{4}\mu^{2}\pm  X_{4}^{\frac{h-1}{h}}\mu^{3}\Bigr]
\end{equation}
where $X_4 = X_4(\theta;\Delta(h))$, $c_2$ is real and arbitrary and $h \geq 2$. It is magnetically dominated since it is Poincar\'e magnetic.

\item Axisymmetric, magnetic:
\begin{equation} 
\int dh A^{M}_{(h, 0)}= \int dh \Phi^h X_1 \mu^2+c.c.
\end{equation}
where $X_1=X_1(\theta,\Delta(h))$ and $h$ is complex. For $h$ real we observed that the spectrum of $h$ is discrete and the lowest value is greater than 4 in appendix.

\item Axisymmetric, magnetic, nontoroidal:
\begin{equation}
A^{NT}_{(h, 0)}=\Phi^{h}X_{2}\Bigl[h \mu^{1} \pm\sqrt{c_{1}}\mu^{3}\Bigr] \label{eq:12bb}
\end{equation}
where $X_2=X_2(\theta,\Delta(h), c_1)$, $h$ is real. After a numerical check involving $X_2$, it turns out that for all $c_1>0$ there exists a range of $1 \leq h \leq h_{max}(c_1)$ where the solution is magnetically dominated for all values of $\theta$. The function $h_{max}$ tends to $1$ in the limit $c_1 \rightarrow 0$ and tends to infinity in the limit $c_1 \rightarrow \infty$. 
It is a solution with no toroidal electric field and no poloidal magnetic field. 

\item Nonaxisymmetric, null:
\bea
\sum_{q \in \mathbb Z} A^{EM}_{(h=1 \pm i k q, q)}&= &\sum_{q \in \mathbb Z} \Phi^{h}\lambda^{q}\Bigl[ha_1(\theta)\mu^{1}\pm {i}qa_1(\theta)\mu^{2}\pm\gamma a_1'(\theta)\mu^3\Bigr]+c.c.\label{Solna1}
\eea
Here $a_1(\theta)$ can be complex. We require that $a_1(\theta)$ and $a_1'(\theta)$ vanish at the poles. 

\item Axisymmetric, null
\bea
A^{EM}_{(1,0)} = \Phi (a_1(\theta) \mu^1 \pm \gamma a_1'(\theta) \mu^3) 
\eea
where $a_1(\theta)$ and $a_1'(\theta)$ vanish at the poles but $a_1(\theta)$ is otherwise arbitrary. This is a special case of the solution \eqref{Solna1} for $q=0$.

\end{itemize}

\subsection{List of potentially physical solutions}
\label{listphys}

In the following we list the real, magnetically dominated or null force-free solutions with nontrivial current and $\text{Re}(h)= \frac{1}{2}$ or $\frac{1}{2} < \text{Re}(h)\leq 1$ and $J(\theta) = 0$ as defined in \eqref{JJdef}, which lead to finite asymptotically flat energy and angular momentum fluxes according to the analysis in Sec. \ref{sec:E} and in \cite{Gralla:2016jfc}. We also do consider linear superpositions. 

\begin{itemize}

\item Nonaxisymmetric, magnetic\\
We have two classes of solutions from Eq. \eqref{solX5real}. The first one for $h(\eta)=\frac{1}{2}+ i\eta $, $\eta \in \mathbb{R}$:
\begin{equation}
A=\int d\eta \sum_{q \in \mathbb Z}\left\{  \Phi^{h(\eta)}\lambda^{q}\left[ X_{5}\mu^{2}-\frac{ i q\gamma(1-k^{2}\gamma^{2})}{q^{2}-\Delta(h(\eta),q)\gamma^{2}} X'_{5}\mu^{3} \right] +c.c. \right\}
\end{equation}
where $X_5=X_5(\theta; h(\eta) ,q)$ and $\Delta(h(\eta),q)= k^2 q^2-\eta^2-\frac{1}{4}$. When $q=0$, the solution reduces to \eqref{solAX12}.

When $\text{Re}(h) \neq \frac{1}{2}$, the constraint $J(\theta) =0$ imposes $h=1$. The solution is
\begin{equation}
A= \sum_{q \in \mathbb Z}\left\{  \Phi \lambda^{q} \left[ X_{2}\mu^{2}-\frac{ i \gamma }{q} X'_{2}\mu^{3} \right] +c.c. \right\}
\end{equation}
after using \eqref{specX5}, where $X_2=X_2(\theta; k^2 q^2 ,q^2)$. 

\item Nonaxisymmetric, null

The solution \eqref{eq56} has $J(\theta)=0$ for $\mu = \pm k q$. This leads to the null solution
\begin{equation}\label{X2a}
A_{(q)}=\Phi^{1\pm i k q}\lambda^{q}\Bigl[(i k q \pm 1)X_2 \mu^1+ iq X_2\mu^2+\gamma X_2'\mu^3\Bigr] +c.c.
\end{equation}
where $X_2=X_2(\theta;\pm i k q, c_1=q^2)$.

\item Axisymmetric, magnetic
\begin{equation}\label{solAX12}
A = \int dh \Phi^{h} X_{1} \mu^{2} + c.c.
\end{equation}
where  $X_1=X_{1}(\theta ; \Delta(h))$ and $h$ is complex in the range $\frac{1}{2} \leq \text{Re}(h)\leq 1$. It is not clear whether regular solutions exist in that range. Indeed, at least for $h$ real, the spectrum of $h$ is discrete for regular solutions and the lowest value is greater than 4, see Appendix A. Since the solution is axisymmetric, we can check the Znajek's condition 
\bea\label{Zn}
I(\psi) = (\Omega(\psi)-\Omega_H) \p_\theta \psi \sqrt{\frac{g_{\phi\phi}}{g_{\theta\theta}}}.
\eea
After taking the near-horizon limit, the generator of the black hole horizon is $\p_t$ so the angular velocity at the Poincar\'e horizon of the near-horizon geometry is $\Omega_H=0$. We also have $I(\psi)=\Omega(\psi) = 0$ as shown in \eqref{eqpsi11}, and therefore \eqref{Zn} holds.
The second regularity condition that should be obeyed for extremal black holes only, as described in \cite{Gralla:2014yja}, is also trivially satisfied.
The solution is also regular in the interior upon choosing global generators \eqref{c11} as seen from \eqref{phi11}.

\item Axisymmetric, magnetic \\
The solution  \eqref{eq:12bb} has $J(\theta) = 0$. For $h=1$, it is magnetically dominated and therefore admissible. It reads as
\begin{equation}
A=\Phi X_{2}\Bigl[ \mu^{1} \pm\sqrt{c_{1}}\mu^{3}\Bigr] 
\end{equation}
where $X_2=X_2(\theta,0, c_1)$ and $c_1 > 0$. Znajek's condition does not apply because the field is nontoroidal, $i_{\p_\phi}F=0$. 

\item Nonaxisymmetric, null \\
Solutions \eqref{Solna1} with $h(q)=1\pm i k q$ have $J(\theta) = 0$:
\begin{equation}
A= \sum_{q \in \mathbb Z}\Phi^{h(q)}\lambda^{q}\Bigl[h(q)a_1(\theta)\mu^{1}\pm {i}qa_1(\theta)\mu^{2}\pm\gamma a_1'(\theta)\mu^3\Bigr]+c.c 
\end{equation}
We require $a_1$ to vanish at the poles. For $q=0$, we obtain the axisymmetric null solution:
\item Axisymmetric, null 
\bea
A_{(1,0)} = \Phi (a_1(\theta) \mu^1 \pm \gamma a_1'(\theta) \mu^3) .\label{solLR}
\eea

\end{itemize}

As discussed in Appendix \ref{app:ODE}, solutions to the ODEs for $X_1$, $X_2$ and $X_{5}$ exist which are regular at the north and south poles. However, the functions $X_1$ and $X_5$ are generically logarithmically divergent at the velocity of light surface. Since the fate of the velocity of light surface is unclear when extending these solutions to the asymptotically flat region, we do not exclude them and consider them as potentially physical. A more complete analysis of the extension of these solutions to the asymptotically flat region would however be necessary to fully settle the issue. 

Generically, highest weight solutions with non-zero weight have a singular field strength at the Poincar\'e horizon. However, the energy and angular momentum fluxes are regular at the Poincar\'e horizon as a consequence of the restriction \eqref{boundh}.  One exception is the positive branch of \eqref{solLR} which is up to a gauge transformation $A _{(1,0)}=a_1(\theta) d (t-\frac{1}{r})$. This solution, found originally in \cite{Lupsasca:2014hua}, is regular at the future horizon but singular at the past horizon. 

The solutions are written in a $SL(2,\mathbb R)$ covariant manner and one can choose any $SL(2,\mathbb R)$ generators related by isomorphisms of the algebra, as discussed around \eqref{auto}.

\section{Summary and conclusion}

This paper provides new analytical solutions to force-free electrodynamics in the near-horizon region of extremal Kerr black holes. We organized and classified them according to the highest-weight representation labeled by the complex weight $h$ and the integer charge (or angular momentum label) $q$. Our classification refines and extends the previous results in the literature. Since the force-free equations are highly nonlinear, we cannot claim to have fully solved them all. Nevertheless, we were able to organize the solutions into seven independent and complete classes. Besides such a mathematical classification, we gave a more physical description of the electromagnetic properties of the solutions: we scrutinized the linear superpositions leading to real magnetically dominated or null solutions which admit finite energy and angular momentum with respect to the asymptotically flat observer. We ended with several families of solutions which, according to our mentioned criteria, are the only potentially physical solutions among all the ones obtained. However, all of them, except one, are singular at the future Poincar\'e horizon.

The second main result of this paper is to have developed the consequences of the fact that the region beyond the velocity of light surface is not physical, because of the lack of a timelike Killing vector field and the consequent impossibility to define a vacuum for the electron-positron plasma in that region. We discussed that it implies that the region near the equator of the near-horizon geometry is unphysical and for this reason we only required that physical solutions be magnetically dominated or null in the polar regions, namely, between the poles and the velocity of light surface. Our third result is more technical and consists in solving numerically the three linear ODEs which appear in highest-weight solutions. We showed that they are regular at the poles but pointed to logarithmic divergences at the velocity of light surface. 

The main open question left is how to glue the near-horizon solutions to asymptotically flat spacetime. We hope to return to this question in the near future.

\section*{Acknowledgments}

G.C. would like to thank A. Strominger, A. Lupsasca and S. Gralla for all their input, encouragement and interesting discussions around this topic that took place at Harvard at the Center for Fundamental Sciences. G.C. would like to thank S. Sheikh-Jabbari for collaboration at the early stages of this project. G.C. is a Research Associate of the Fonds de la Recherche Scientifique F.R.S.-FNRS (Belgium). G.C. and R.O. acknowledge the current support of the ERC Starting Grant No. 335146 ``HoloBHC." This work is also partially supported by FNRS-Belgium (convention IISN 4.4503.15).

\appendix

\section{Relevant ordinary differential equations}
\label{app:ODE}
In this section we list the five ordinary differential equations whose solutions are present in Sec. \ref{allsol}. 

During the resolution of the three coupled nonlinear ODEs in $\theta$ as described in Sec. \ref{sec:solv},  we encountered the following nonlinear ODE for $X(\theta ; h,c_1,c_2)$:
\bea \label{DiffX}
&h^2 X^{2}\Bigl[X(\gamma'X'-\gamma X'')-\gamma(h-1)(c_2^2+X^2+X'^2)\Bigr]+ \gamma^2  \gamma' X X' \Bigl[(h-1)c_1 + h k X\Bigr]^2 \nonumber \\
&+\gamma^3 \Bigl[(h-1)c_1 + h k X\Bigr] \Bigl\{(h-1)\Bigl[h X^2 \left(c_1+kX\right)+\bigl[(h-2)c_1 +hkX\bigr] X'^2\Bigr]\nonumber \\
&\hspace{3.8cm} +X\Bigl[(h-1)c_1 + h k X\Bigr]X''\Bigr\} =0.
\eea
In the case $c_1 \neq 0$, one might substitute $X =c_1 X_3$ and define $\xi \equiv \frac{c_2^2}{c_1^2}$. Then all the dependence in $c_1$ factors out. We then obtain the differential equation for $X_3(\theta ; h, \xi )$ which is listed below. This ODE was also found in \cite{Zhang:2014pla}. Upon setting $c_1 = 0$ one gets another nonlinear ODE. When $c_2 \neq 0$, $h \neq 0$, $c_2$ can be factored out of the equation upon a rescaling of $X$. We denote the resulting function as $X=c_2 X^{1/h}_4(\theta ; \Delta(h))$. The ODE for $X_4$ is listed below. When both $c_1=c_2=0$ and $h \neq 0$ we find a linear ODE that we denote as $X=X^{1/h}_1(\theta ; \Delta(h))$. 

Recall that the unphysical region beyond the velocity of light surface lies in the range $\theta_* \leq \theta \leq \pi - \theta_*$. In the following, we will assume that all functions $X(\theta)$ together with their first derivatives are finite in the physical region $0 \leq \theta \leq \theta_*$ and $\pi-\theta_* \leq \theta \leq \pi$, i.e., $X(\theta)< \infty$, $X'(\theta) < \infty$.

For the extremal Kerr black hole $k=1$ and $\gamma,\Gamma$ are given in \eqref{paramKerr}. In particular it is useful to note that $\gamma(\pi-\theta)=\gamma(\theta)$, $\gamma(0)=\gamma(\pi)=0$, $\gamma'(0)=1$, $\gamma'(\pi)=-1$ where $\theta = 0$ is the north pole and $\theta = \pi $ is the south pole. Also, $\theta_*=\arcsin [\sqrt{3}-1] \sim 0.82$ is the lowest positive root of $\gamma(\theta)-1$. All numerical solutions will be plotted only for the extremal Kerr black hole.

\subsection*{1) $X_{1}(\theta ; \Delta(h))$}
\begin{equation}\label{Diff1}
\text{ODE}_1[X_1; \Delta(h)] \equiv X_1''+ \frac{\gamma'}{\gamma} \frac{k^2 \gamma^2+1}{k^2 \gamma^2 - 1} X_1' + \Delta(h)X_1 =0,
\end{equation}
where $\Delta(h)=h(h-1)$. This ODE appears only in solution \eqref{eq: 11}. From the physical requirement that there should not be any singular magnetic flux at the north and south poles; see \eqref{eqpsi11}: we impose the boundary condition $X_1(0) =X_1(\pi) = 0$. 

The equation is invariant under the transformations $h \rightarrow 1-h$ which leaves $\Delta(h)$ invariant: that is $X_{1}(\theta; h)=X_{1}(\theta; 1-h)$. Moreover, the equation is invariant under the reflection $\theta \rightarrow \pi-\theta$ so that if $X_1(\theta)$ is a solution, so is $X_1(\pi - \theta)$.\footnote{The function solution to this ODE was denoted as $S_h(\theta)$ or $S_{h,m=0}(\theta)$ in \cite{Lupsasca:2014hua} but was not explicitly solved.}

For the special case $\Delta=0$, i.e. $h=0$ or $h=1$, the solution is 
\bea
X_1(\theta ; 0) = C_1 + C_2 \frac{5\sin\theta+\sin 3\theta}{19-16k^2 +4(3+4k^2) \cos 2\theta+\cos 4 \theta}.
\eea
However, regularity at the velocity of light surface implies $C_2=0$ and $X_1(\theta,0)$ is therefore constant which we fix to 0 by the boundary condition $X_1(0)=0$. There is therefore no solution. 

In general, the differential equations have regular singular points at the zeros of $\gamma$ and $k^2\gamma^2 - 1$ which are located at $\theta_o=0,\theta_*,\pi-\theta_*,\pi$. Indeed, for each root $\theta_o$ we have 
\bea
 \frac{\gamma'}{\gamma} \frac{\gamma^2+1}{\gamma^2 - 1}=\frac{1}{\theta-\theta_{o}}+ \mbox{regular terms}.
\eea
Therefore, Frobenius' method is applicable. In the generic case, the solution reads close to the pole $\theta_o$ as a linear superposition of the power series solutions $(\theta-\theta_{o})^{\lambda_\pm}\sum_{n=0}a_n(\theta-\theta_{o})^n$ where $\lambda_\pm$ are the two roots of the indicial equation. In case of double roots, a logarithmic branch appears. Frobenius' series converges in the open complex disk that contains only one root.

In the range $0 \leq \theta \leq \theta_*$ we could start the series from the north pole or the velocity of light surface. Now, the indicial roots are 0 and 2 around the north pole while there is a double root 0 around the velocity of light surface which leads to a logarithmically divergent solution. One might however question whether such a logarithmic divergence is admissible. After all, the geometry around the velocity of light surface will be modified significantly when considering the asymptotically flat extension of the geometry.  If the logarithmic divergence at the velocity of light is acceptable, one can simply write an expansion close to the north pole $X_1 \sim a_0 \theta^0 + a_2 \theta^2 +O(\theta^4)$ with the boundary condition $a_0=1$ and the choice of normalization $a_2=1$. The solution then exists for all (complex) values of $h$ and is defined by a power series expansion. 

\begin{figure}[!htb] 
\centering
\includegraphics[width=12cm, height =6cm]{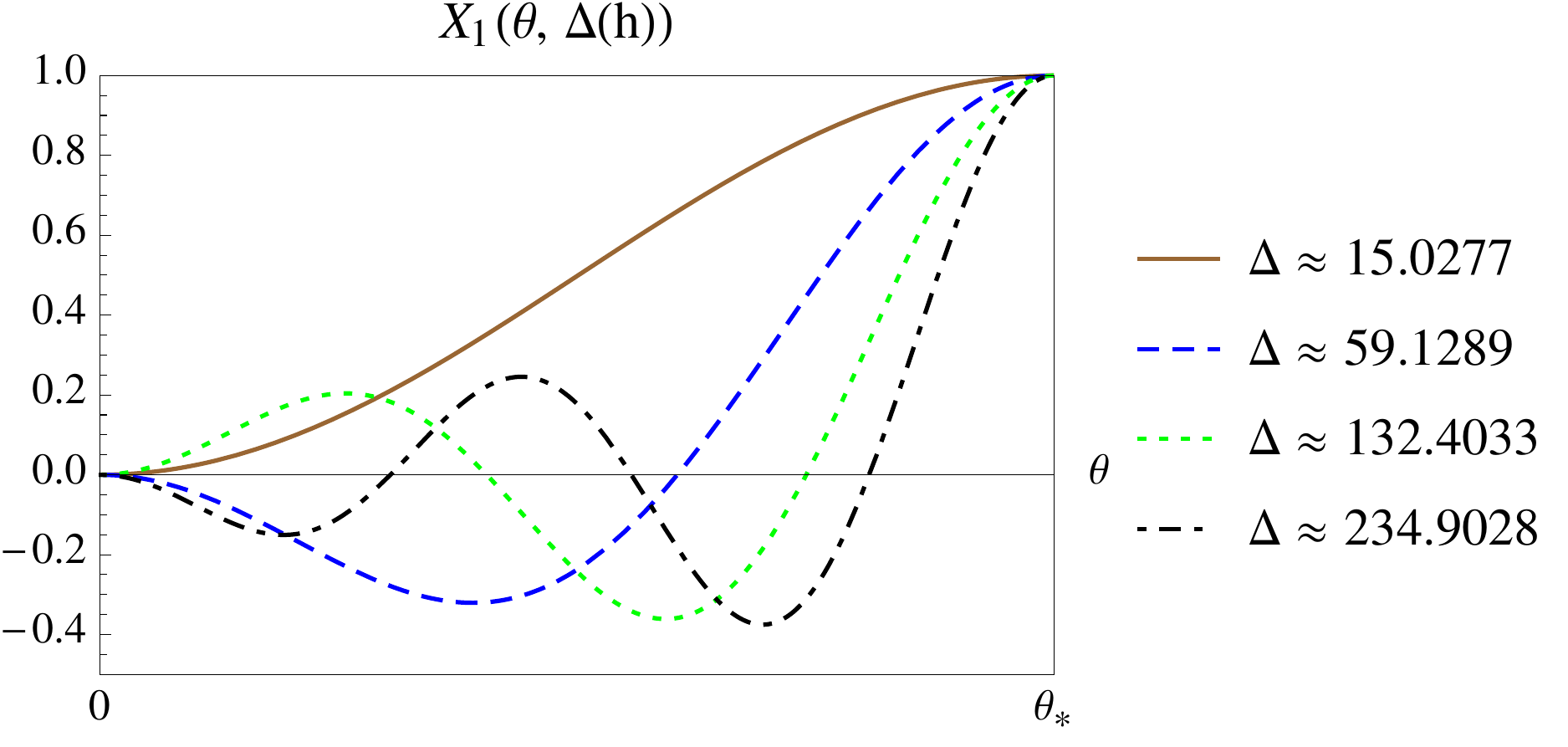}
\caption{\small Solutions to $\text{ODE}_1[X_1; \Delta(h)]$ for the first discrete values of the parameter $\Delta(h)$ which obey the boundary conditions $X_1(0)=0$, $X_1(\theta_*)=1$.}
\label{fig: ODE1}
\end{figure}

If one removes the logarithmic branch, there is only one constant of integration which we fix by setting $X_1(\theta_*; \Delta(h)) = 1$. The solution then takes the form 
\bea
X_1(\theta; \Delta(h))=1+\sum_{n\geq1}a_n(h)(\theta-\theta_{*})^n
\eea
where the coefficients $a_n(h)$ can easily be obtained. Now, the physical boundary condition $X_1(0)=0$ will be obtained only for a discrete spectrum of $\Delta(h)$. The first real values of $\Delta(h)$ together with the plot of the solution $X_1(\theta ; \Delta(h))$ in the physical domain are given in Fig. \ref{fig: ODE1}.

\subsection*{2) $X_{2}(\theta; \Delta(h,q),c_{1})$}
\begin{equation}\label{Diff2}
\text{ODE}_2[X_2 ; \Delta(h, q), c_1] \equiv X_2'' + \frac{\gamma'}{\gamma} X_2' +  \Bigl[ \Delta(h,q) -\frac{c_1}{\gamma^2} \Bigr]X_2 =0
\end{equation}
where $\Delta(h,q) = h (h-1)  + k^2q^2$. This equation appears in \eqref{X2a} with $c_1 = q^2$, in \eqref{eq:12} with $c_1$ arbitrary and in \eqref{eq: AD1} with $c_1=0$. The equation is invariant under all transformations of $h,q$ that leave $\Delta(h, q)$ invariant.  
In the axisymmetric case $q=0$ and when $c_1 \neq 0$ we require from the analysis around \eqref{eq:r1} the boundary condition $X_2(0)=X_2(\pi)=0$.

\begin{figure}[htb] 
\centering
\includegraphics[width=12cm, height =6cm]{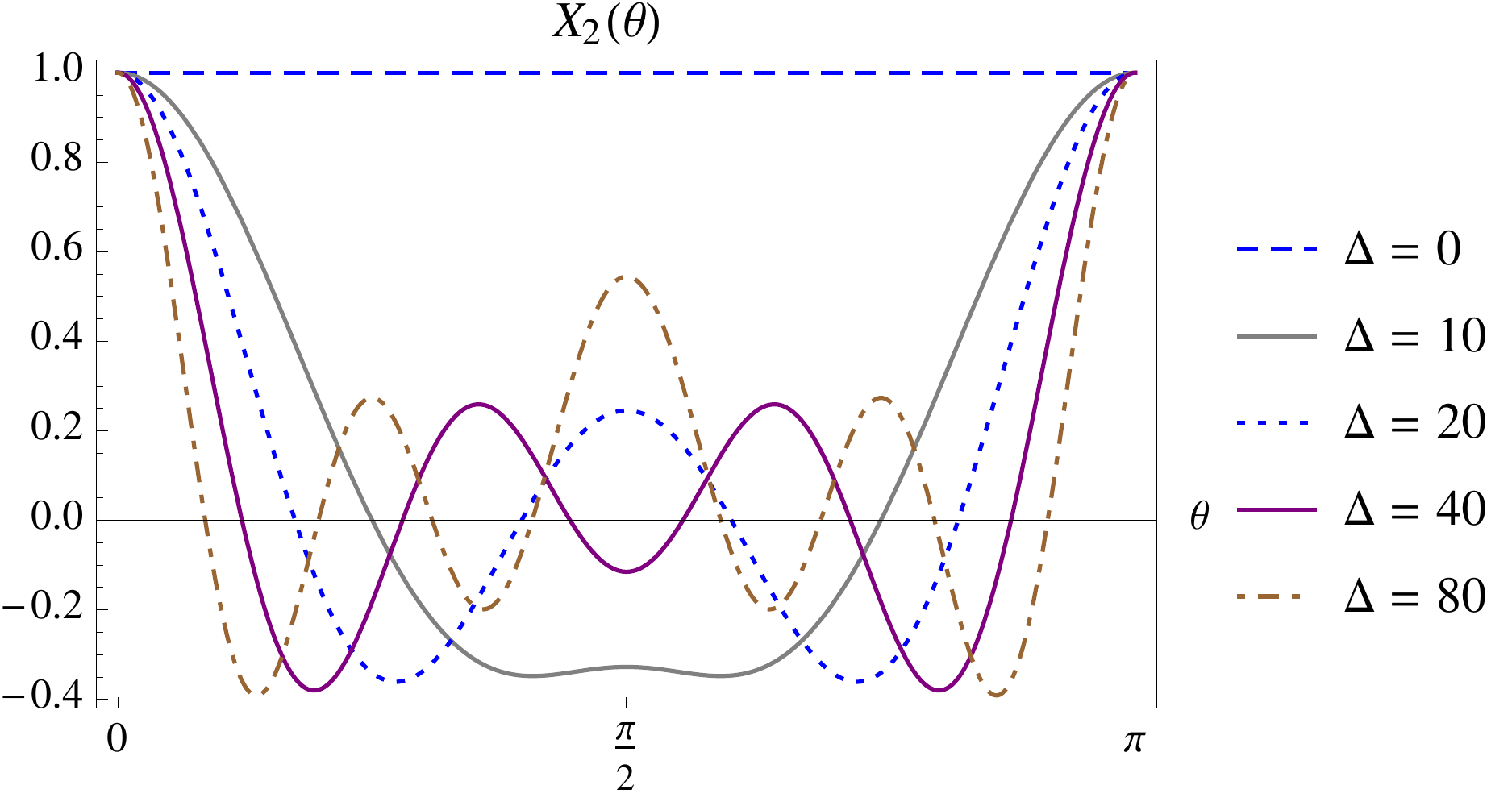}
\caption{Solutions to $\text{ODE}_2[X_2; \Delta(h,q); 0]$ for different values of the parameter $\Delta(h,q)$.}
\label{fig: ODE2c1=0}
\end{figure}

Let us first analyze the simplest case $c_1=0$. This $\text{ODE}_2$ generalizes the one written in \cite{Lupsasca:2014hua} [see their eq. (4.9)] for $q$ arbitrary. We concentrate on solutions which are symmetrical around the equator and therefore we only consider the interval $\theta = [0,\frac{\pi}{2}]$. After performing the substitution $x=\sin^{2}(\theta)$, $\text{ODE}_2$ takes the form of the generalized Heun's equation 
\bea\nonumber
X''_{2} + \frac{(\alpha+\beta +1)x^{2}-(\alpha+\beta +1+a(\gamma+\delta)-\delta)x+a\gamma}{x(x-1)(x-a)}X'_{2} +\frac{\alpha\beta x -b}{x(x-1)(x-a)}X_{2}=0
\eea
where
\bea
a=2, \quad b=-\frac{\Delta}{2}, \quad \alpha\beta=-\frac{\Delta}{4}, \quad \alpha+\beta=-\frac{1}{2}, \quad \gamma=1, \quad \delta=\frac{1}{2}.
\eea
Frobenius' method can be applied. There are poles at the north and south poles and at the fake pole $x=2$. One could therefore expand in a power series at the north pole and it will converge over the region $x \in [0,1[$. It is important to note that the equator $x=1$ ($\theta = \frac{\pi}{2}$) is not included in the radius of convergence so care should be taken. At the north pole we find 0 as a double root of the indicial equation. Removing the logarithmically divergent branch, we get the following regular convergent power series in the domain $\theta \in [0, \frac{\pi}{2}[$:
\bea \label{SolOde2}
X_{2}(\theta; \Delta(h,q))=\sum_{n=0}^{\infty}d_{n}(\Delta(h,q))\sin^{2n}(\theta)
\eea
The solution \eqref{SolOde2} behaves close to the north pole as $X_{2}(0)=d_0$ where $d_0$ is arbitrary (which we fix to 1 by linearity of the equation) and $X_2'(0)=0$. The coefficients obey the second-order recurrence relation $d_{n+1}=A_{n}d_{n}+B_{n}d_{n-1}$ with $d_{1}=-\frac{\Delta}{4}d_0$ and 
\bea
A_{n}=\frac{6n^{2}-\Delta}{4(n+1)^{2}}, \quad B_{n}=-\frac{2(n-1)(2n-3)-\Delta}{8(n+1)^{2}}.
\eea
Now, a numerical convergence analysis reveals that the series expansion does not converge at the equator $x=1$ unless $\Delta = 0$. We interpret this by the presence of a source at the equator for generic values of $\Delta$. In Fig. \ref{fig: ODE2c1=0}, we plot $X_2(\theta, \Delta(h, q))$ obtained from the series expansion truncated to order 20 for some real values of the parameter $\Delta(h,q)$ (the value of the function around $\frac{\pi}{2}$ for $\Delta \neq 0$ should be taken with a grain of salt since the series does not converge there).

Let us now analyze the more general case $c_{1} \neq 0$. \footnote{In the case $c_1=q^2$ this equation was considered in \cite{Lupsasca:2014hua}; see their (5.23).} The $\text{ODE}_2$ is still only singular at the zeros of $\gamma$, i.e., at points $\theta=0, \pi$. After performing the substitution $z=\sin^2(\theta)$, the $\text{ODE}_2$ takes the following form
\begin{equation}
X''_2(z) +\frac{z^2-6z+4}{2z(z-1)(z-2)}X'_2(z) +\frac{(z-2)^2 c_1 -4\Delta z}{16z(z-1)}X_2(z)=0
\end{equation}
This makes it clear that $z=0,1$ are regular singular points so that Frobenius's method applies. The two solutions of the indicial equation at the north pole are $\lambda_{\pm}=\pm \frac{\sqrt{c_1}}{2}$. In order to avoid oscillations at the poles we enforce $c_1 > 0$ from now on. Only the solution $\lambda_+$ is admissible since otherwise the solution will diverge at the north pole. In the special case where $\lambda_+ - \lambda_-$ is an integer $q$ which incidentally occurs for $c_1=q^2$, one independent solution contains a logarithmic branch which again diverges. Again in this case, only the solution which behaves as $z^{\lambda_+}$ is admissible. 

\begin{figure}[htb] 
\centering
\includegraphics[width=12cm, height =6cm]{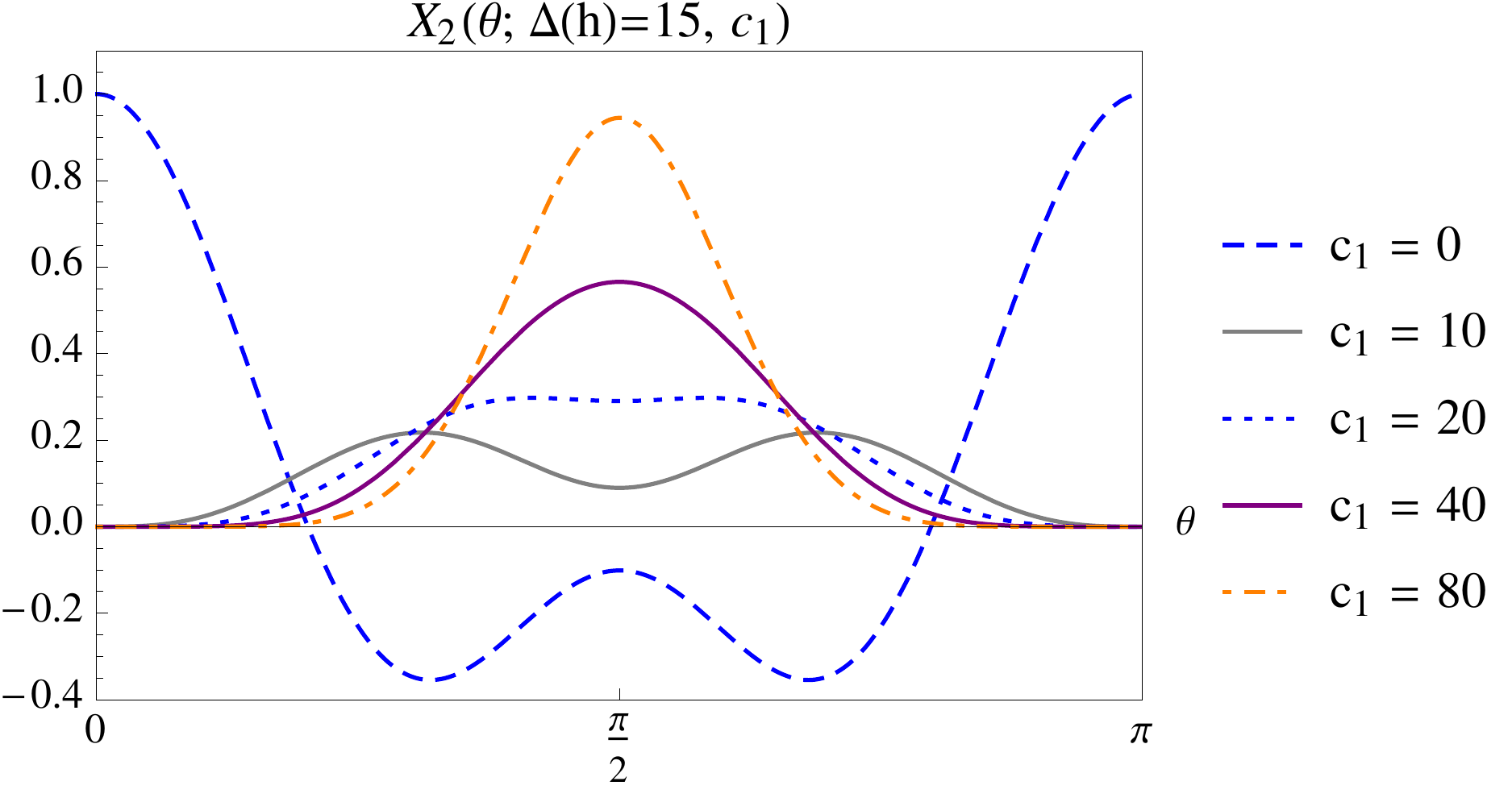}
\caption{Solutions to $\text{ODE}_2[X_2 ; \Delta(h,q), c_1]$ for $\Delta(h,q) =15 $ and different values of $c_1$.}
\label{fig: ODE2c1_1}
\end{figure}

In all cases, the regular solution is given for $c_1 > 0$ by
\bea\label{solX2}
X_2(z) = z^{\lambda_+} \sum_{n=0}^{\infty} a_n z^n 
\eea
where $a_0$ is an arbitrary constant and 
\begin{align}
a_1&=-\frac{\Delta}{4(1+\sqrt{c_1})}a_0\\
a_2&=\frac{\Delta^2 -4\Delta(1+ \sqrt{c_1}) + 3c_1 + (2+c_1)\sqrt{c_1}}{32(2 +c_1 + 3 \sqrt{c_1})}a_0
\end{align}
For $n \geq 2$, we have $a_{m+1}=A_{m}a_{m}+B_{m}a_{m-1}+C_{m}a_{m-2}$ where
\begin{align}
A_{m}&= \frac{6m(m + \sqrt{c_1})-\Delta}{4(m+1)(m+1 + \sqrt{c_1})},\\
B_{m}&= \frac{2\Delta -4(m-1)(2m-3) +c_1 + 2(9-4m)\sqrt{c_1}}{16(m+1)(m+1 + \sqrt{c_1})},\\
C_{m}&= -\frac{c_1}{32(m+1)(m+1+ \sqrt{c_1})}.
\end{align}
We fix $a_0=1$ without loss of generality. In the case $c_1=0$, we recover the recurrence relation \eqref{SolOde2}. From \eqref{solX2} we directly see that for all $c_1 > 0$, the function $X_2$ obeys the boundary condition $X_2(0) = X_2(\pi) = 0$. We again observe numerically that the series \eqref{solX2} does not converge at the equator $\theta = \frac{\pi}{2}$ unless $c_1$ is fixed as a definite function of $\Delta(h)$,
\bea
c_1 = c_1(\Delta(h))
\eea
which asymptotes to 0 for $\Delta=0$ and to $\infty$ for $\Delta \rightarrow \infty$. For example, for $c_1(10) \approx 4.90$, $c_1(20) \approx 8.48$, $c_1(40) \approx 15.02$, $c_1(80) \approx 27.23$. 
In Fig. \ref{fig: ODE2c1_1}, we plotted the power series solution truncated to order 20 to $\text{ODE}_2[X_2 ; \Delta(h,q) = 15, c_1]$ for different values of the parameter $c_1$. Note that the boundary condition $X_2(0)=0$ is true only for $c_1>0$. 

\begin{figure}[htb] 
\centering
\includegraphics[width=12cm, height =6cm]{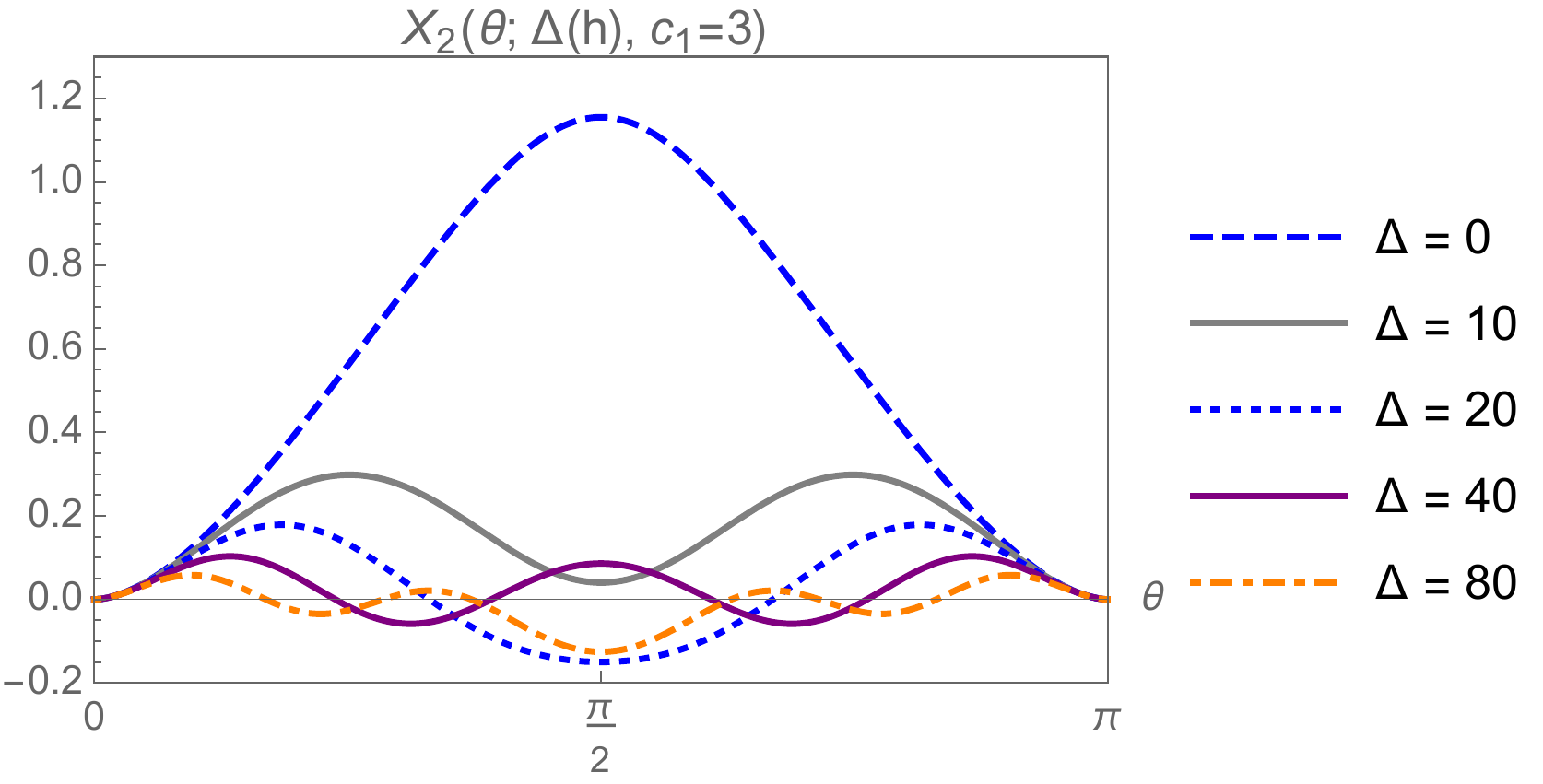}
\caption{Solutions to $\text{ODE}_2[X_2 ; \Delta(h,q); c_1]$ for $c_1=3 $ and different values of $\Delta(h,q)$.}
\label{fig: ODE2c1_2}
\end{figure}

In Fig. \ref{fig: ODE2c1_2}, we studied the behavior of solutions (except at $\theta = \frac{\pi}{2}$) to $\text{ODE}_2[X_2 ; \Delta(h,q) , c_1=3]$ by varying $\Delta(h,q)$ at constant $c_1=3$.

\subsection*{3) $X_{3}(\theta; h,\xi)$}

\begin{equation}\label{Diff3}
\begin{split}
\text{ODE}_3[X_3 ; h, \xi] \equiv & \; h^2 X_3^{2}\Bigl[X_3(\gamma'X_3'-\gamma X_3'')-\gamma(h-1)(\xi+X_3^2+X_3'^2)\Bigr] \nonumber \\
&\hspace{-2cm}+ \gamma^2  \gamma' X_3 X_3' \Bigl[(h-1)+ h k X_3\Bigr]^2 
+\gamma^3 \Bigl[(h-1) + h k X_3\Bigr] \Bigl\{(h-1)\Bigl[h X_3^2 \left(1+kX_3\right) \nonumber\\
&+\bigl[(h-2) +hkX_3\bigr] X_3'^2\Bigr]+X_3\Bigl[(h-1) + h k X_3\Bigr]X_3''\Bigr\} =0.
\end{split} 
\end{equation}
This ODE appears in solution \eqref{eq: 3}. According to the discussion around \eqref{eqpsi11cc} we require the boundary condition $X_3(0)=X_3(\pi)=0$. 

When $h=0$, the equation reduces to 
\bea
\p_\theta \left( \gamma \p_\theta (X_3^{-1})\right) = 0.
\eea
When $h=1$, the equation reduces to $\text{ODE}_1[X_3(\theta),\Delta=0]$, therefore 
\be
X_3(\theta,h=1,\xi) =   X_1(\theta ; \Delta=0).
\ee
We will not solve this nonlinear ODE here.  When $h \neq 0,1$, the equation was considered in \cite{Zhang:2014pla,Lupsasca:2014hua} and solved  in \cite{Zhang:2014pla} for the case $h=-1$.

\subsection*{4) $X_{4}(\theta; \Delta(h) )$}

\begin{equation}\label{Diff4}
\begin{split}
\text{ODE}_4[X_4 ; \Delta(h)] \equiv  X_4''+ \frac{\gamma'}{\gamma} \frac{k^2 \gamma^2+1}{k^2 \gamma^2 - 1} X_4' + \Delta(h)X_4 +\frac{\Delta(h) }{1-k^2 \gamma^2}X_4^{\frac{h-2}{h}} = 0
\end{split} 
\end{equation}
where $\Delta(h)=h(h-1)$ which we can rewrite as
\bea \label{X4}
\text{ODE}_1[X_4 ; \Delta(h)] +\frac{\Delta(h) }{1-k^2 \gamma^2}X_4^{\frac{h-2}{h}} = 0. 
\eea
This ODE appears in solution \eqref{eq: 3b}. According to the discussion around \eqref{eqpsi112} we require the boundary condition $X_4(0)=X_4(\pi)=0$. 

We will not solve this nonlinear ODE here. Note that when $h =2$ the equation becomes linear with a nonhomogenous term.

\subsection*{5) $X_{5}(\theta ; h,q)$}
\begin{equation}\label{Diff5}
\begin{split}
X_5''+\frac{\gamma'}{\gamma}  \Bigl[1 - \frac{2 h (h-1) \gamma^2}{(1-k^2 \gamma^2)(\Delta(h,q) \gamma^2-q^2)}\Bigr]X_5'+ \left[  \Delta(h,q) - \frac{q^2}{\gamma^2} \right]X_5 = 0
\end{split} 
\end{equation}
where $\Delta(h,q)=h(h-1)+k^2 q^2$. This linear ODE appears in solution \eqref{solX5}.

 It admits the symmetries 
\bea
X_{5}(\theta ; h,q)=X_{5}(\theta ; 1-h,q)=X_{5}(\theta ; h,-q)=X_{5}(\theta ; 1-h,-q).
\eea 
We note the special cases 
\bea
X_5(\theta ; h ,q=0) &=& X_1(\theta , \Delta(h) ),\\
X_5(\theta ; h=0 , q) &=&X_2(\theta ; \Delta = k^2 q^2 , q^2) ,\\ 
X_5(\theta ; h=1 , q) &=&X_2(\theta ; \Delta = k^2 q^2 , q^2).\label{specX5}
\eea
Since the ODE for $X_1$ was analyzed previously we concentrate on $q \neq 0 $ only. \footnote{This ODE was also found in \cite{Lupsasca:2014hua}; see their (3.29) where $X_5$ is denoted as $S_{h,m}$. Our analysis of the ODE however slightly differs.}

There are always two regular singular points in the range $0 \leq \theta \leq \frac{\pi}{2}$: first at $\theta = 0$ (north pole), and then at $\theta = \theta_* =  \arcsin(\sqrt{3}-1) $ (velocity of light surface). When $h$ is real and for $q \neq 0$, there is a regular singular point at the real root of $\Delta \gamma^2 = q^2$ which is always in the range $0 \leq \theta \leq \frac{\pi}{2}$. There is also an imaginary root which obeys $\sin\theta \geq \sqrt{\frac{1}{2}(7+\sqrt{33})} \approx 2.52$ (bound reached at $h=\frac{1}{2}$, $q=1$) so it is irrelevant for discussing convergence in the interval $0 \leq \sin\theta \leq 1$. 

\begin{figure}[!hbt] 
\centering
\includegraphics[width=12cm, height =6cm]{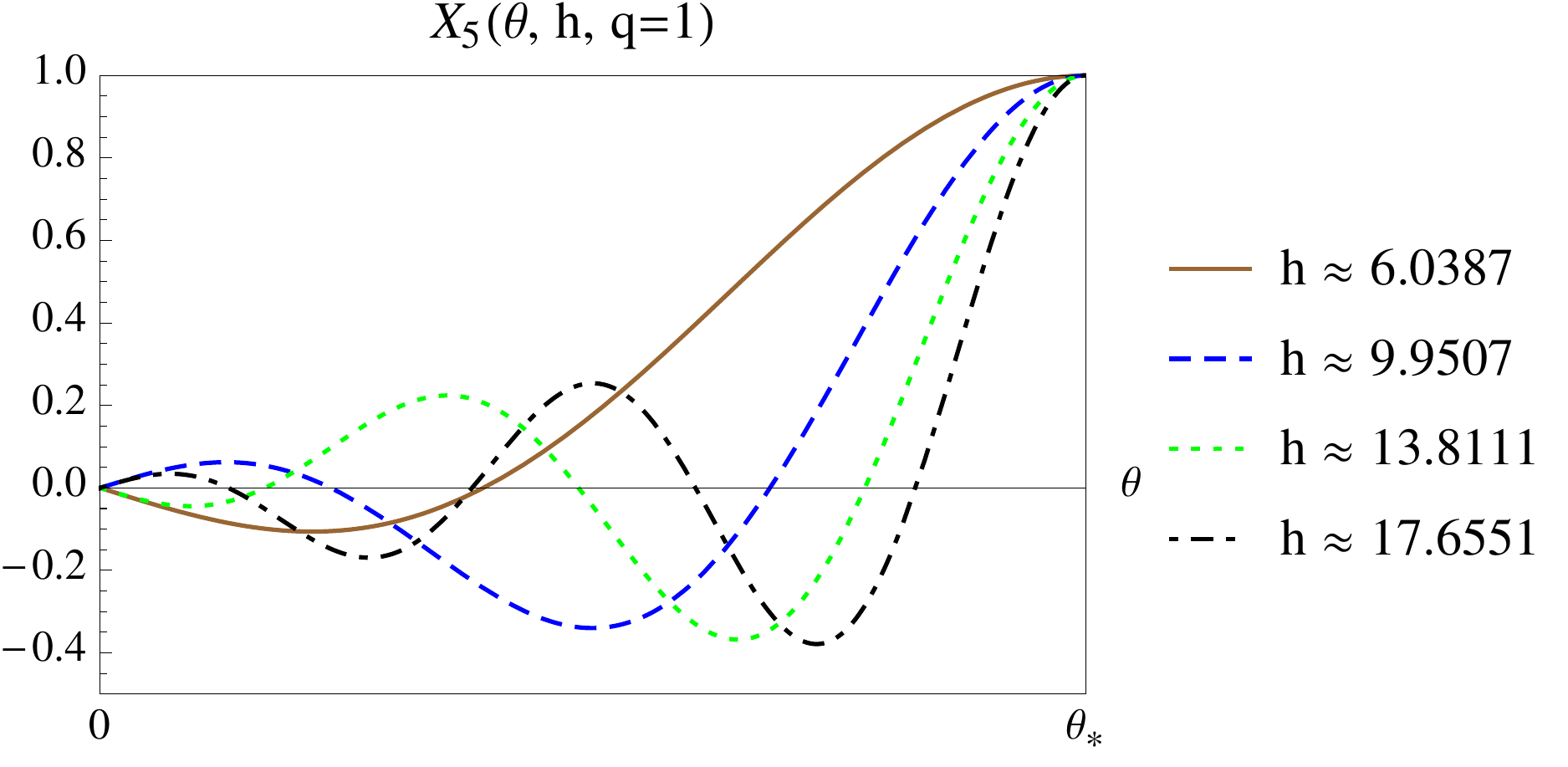}
\caption{Solutions to $\text{ODE}_5[X_5; h; q=1]$ for $q=1 $ and different values of $h$ such that $X_5(0) = 0$. }
\label{fig: ODE5}
\end{figure}

The two independent solutions behave close to $\theta = 0$ as $\theta^q$ and $\theta^{-q}$, while they behave close to $\theta = \theta_* $ as $\log( \theta-\theta_*)$ and $(\theta-\theta_*)^0$. If one only insists in having a solution smooth at the north (and south) poles, a solution always exists but it will be generically logarithmically divergent at the velocity of light surface. In order to avoid singularities, we need to interpolate between the solutions $\theta^q$ (we assume $q >0$) at $\theta = 0$ and $(\theta-\theta_*)^0$ at $ \theta = \arcsin(\sqrt{3}-1)$. This involves a shooting method which will discretize the possible values of $h$ as a function of $q$. We then normalize the solution with $X_5(\theta_*)=1$. There is therefore no more free continuous constant of integration. In the range $0 \leq h \leq 1$ and $q >0$ the other singularities are not in the range between the north pole and the velocity of light surface. For $h \geq 1$ and $ h \leq 0$, they are but we checked that the indicial equation around the real pole and imaginary pole has exponents 0 and 2 so an interpolating function between the velocity of light surface and the north pole will be smooth. The regular solutions for the first four real values of $h$ are depicted on Fig. \ref{fig: ODE5} for $q=1$. 

\section{Properties of all highest-weight solutions}
\label{ID}

In this section, we analyze the properties of all solutions listed in Sec. \ref{allsol}. 
In particular, for each vector potential we compute and list its field strength and its current. These properties constitute the ID card of each solution. We also compute the canonical Euler potentials (for definitiveness in Poincar\'e coordinates) defined in Sec. \ref{cep}. We finally check for regularity of the solutions at the poles in order to derive the relevant boundary conditions for the ODEs that the solutions depend on. 

We recall the definitions $\Delta(h)=h(h-1)$ and $\Delta(h,q)=h(h-1)+k^2q^2$.

\subsection*{$(h,q)$-eigenstates}
There are two classes of solutions with arbitrary nonvanishing highest-weight $h$ and $U(1)$-charge $q$, describing stationary and nonaxisymmetric field configurations.

\subsubsection*{Poincar\'e magnetic}
\begin{empheq}[box=\fbox]{align*}
A_{(h, q)}&=\Phi^{h}\lambda^{q}\left[ X_{5}\mu^{2}-\frac{ i q\gamma(1-k^{2}\gamma^{2})}{q^{2}-\Delta(h,q)\gamma^{2}} X'_{5}\mu^{3} \right] \\ 
F_{(h,q)}&=h\Phi^{h-1}\lambda^{q} \Biggl[ \frac{\gamma X'_5}{q^{2}-\Delta(h,q)\gamma^{2}}\Bigr(-iq(1-k^{2}\gamma^{2})w^{1}+(1-h)k\gamma^2 w^{2}- \frac{w^{3}}{k}\Bigl) \nonumber\\
&\qquad  + X_5(w^{4}-w^{5})\Biggr]\\
J_{(h, q)}&=-\Phi^{h}\lambda^{q}hk\Biggl[X_{5} + \frac{2(1-h)\gamma\gamma'}{(1-k^{2}\gamma^{2})(q^{2}-\Delta(h,q)\gamma^{2})}X'_{5} \Biggr]\frac{\Phi H_{+}}{\sqrt{2}\Gamma^2}
\end{empheq}
where $X_5=X_5(\theta;h,q)$. Since the current keeps its direction upon changing $h,q$, one might linearly superpose solutions with different $h,q$. The current $J_{(h, q)}$ and its complex conjugate $(J_{(h, q)})^{*}$ are proportional to each other. Then, we can obtain a real solution by adding up the vector potential $A_{(h, q)}$ and its complex conjugate. 

We do not consider the trivial case $h=0$, because it gives no field strength $F=0$. For $h=1$, $q \neq 0$ the solution exists and the function $X_5$ becomes $X_5(\theta ; 1,q)=X_2(\theta , k^2q^2,q^2)$. In the case $q=0$, $h \neq 0,1$, we recover the Poincar\'e magnetic solution \eqref{eq: 11} after using the identity $X_5(\theta ; h ,0) = X_1(\theta ; h)$. 

In order to get physical insight, it is useful to derive the functional expression of Euler potentials.   We only consider Poincar\'e coordinates and $q \neq 0$. The field strength describes a stationary and $Q_0$-eigenstate configuration with $i_{\p_t}F=0$  and $df=0$ according to the analysis of Sec. \ref{cep}. Therefore it takes the form \eqref{case2} with 
\bea
\boxed{
\psi(r,\theta)=\frac{X_5}{r^h}, \qquad \p_{\theta}\psi_2(r, \theta)=-iq\Biggl(\frac{1-k^2\gamma^2}{q^2-\Delta(h, q) \gamma^2}\Biggr)\frac{X'_5}{X_5}}
\eea
The Euler potential $\psi$ is singular at the Poincar\'e horizon $r=0$, unless $h$ is negative. In the special case $h=1$ and $q\neq0$, we can easily integrate $\psi_2$ and obtain $\psi_2(\theta)=-\frac{i}{q}\ln(X_5)+\mbox{const}$.\\

\subsubsection*{Poincar\'e generic}

\begin{empheq}[box=\fbox]{align*}
A_{(h, q)}&=\Phi^{h}\lambda^{q}\Bigl[h(h-1)X_2\mu^{1}-kq^2 X_2\mu^{2}+{i}kq\gamma X_2'\mu^{3}\Bigr]\\
F_{(h,q)}&=\Phi^{h-1}\lambda^{q} h \Biggl[\gamma X'_2(ikq w^{1} - (h-1) w^{2}) +\\
 &\qquad +X_2\Bigl(- \frac{k^2q^2+(h-1)^2}{k}w^{4}+kq^2w^{5}+i(h-1)q w^{6}\Bigr)\Biggr] \\
 J_{(h, q)}&=\Phi^{h}\lambda^{q}\frac{hk}{\Gamma}X_{2}\Bigl[(h-1)^{2}\gamma^{2}\hat{Q}_0 + kq^{2} \frac{\Phi H_{+}}{\sqrt{2}\Gamma} - {i}q(h-1)\mu^{4}\Bigr]
\end{empheq}
where $X_2=X_2(\theta; \Delta(h,q),c_1=q^2)$.

The case $h=0$ is trivial since $F=0$. The current is not proportional to its complex conjugate, unless for the special cases $q=0$ or $h=1$ or $h=1+ i \mu$ for any real $\mu$. The case $q=0$ coincides with the solution \eqref{eq: AD1} and it will be analyzed below. The case $h=1$ actually coincides with the solution \eqref{solX5} for $h=1$. It was just analyzed in the previous subsection. The third class is an independent real solution.

In the generic case $q \neq 0$, $h \neq 0,1$, the field strength describes a stationary and $Q_0$-eigenstate configuration with $i_{\p_t}F \neq 0$. The canonical Euler potentials in Poincar\'e coordinates can therefore be written as \eqref{EP1} where 
\bea
\boxed{
\chi_1(r,\theta)=h(h-1)\frac{X_{2}}{r^{h-1}}, \qquad \chi_2(r)=-\frac{{i}kq}{h-1}\frac{1}{r}, \qquad \kappa(e^{iq \phi}\chi_1)=0.}
\eea

\subsection*{$(h \neq 0, q=0)$-eigenstates}

\subsubsection*{Poincar\'e generic}
\begin{empheq}[box=\fbox]{align*}
A_{(h, 0)}&=c_1^h \Phi^{h}\Bigl[- X_{3}^{h-1}\mu^{1}+X_{3}^h\mu^{2}\pm \sqrt{\xi}X_{3}^{h-1}\mu^{3}\Bigr], \qquad c_1\neq0\\
F_{(h,0)}&=c_1^h \Phi^{h-1}X^{h-2}_{3}\Biggl[\pm h \sqrt{\xi}X_3 w^{1} + \gamma ((h-1) +hkX_{3})X'_3 w^{2} - \frac{h}{k\gamma}X'_{3}X_3 w^{3}+\\
&\qquad + \Biggl(hX_{3}+ \frac{(h-1)}{k}\Biggr)X_3w^{4}-hX_{3}^2w^{5}\Biggr]\\
J_{(h, 0)}&=c_1^h \Phi^{h}X^{h-2}_{3}\Biggl[(h-1)A(\theta; h, \xi)\hat{Q}_0 - hX_{3}\frac{h\xi/\Gamma-A(\theta; h, \xi)}{\gamma^{2}}\frac{\Phi H_{+}}{\sqrt{2}\Gamma} \\ 
&\qquad \mp \frac{h(h-1)\sqrt{\xi}}{\Gamma}X_3\mu^{3} \pm\frac{h(h-1)\sqrt{\xi}}{\gamma \Gamma}X'_3 \mu^{4}\Biggr]
\end{empheq}
where $A(\theta; h, \xi)$ is given by
\begin{align}
(\Gamma[\gamma^{2}&(h-1+hkX_{3})^{2}-h^{2}X_{3}^{2}])A(\theta; h, \xi)= \nonumber\\
\qquad &X_{3}[h^{2}\xi-\gamma^{2}(h-1+hkX_{3})][-hX_{3}+k\gamma^{2}(h-1+hkX_{3})]\nonumber \\
\qquad &+2h\gamma X_{3}(h-1+hkX_{3})\gamma'X'_{3}-h(h-1)\gamma^{2}X_{3}^{\prime 2}
\end{align}
where $X_3=X_3(\theta ; h,\xi)$. In the case $h, c_1 \in \mathbb{R}$ and $\xi \geq 0$, the expression of the vector potential $A_{(h, 0)}$ is real. 
For $h=1$, we get the Poincar\'e generic solution \eqref{eq: 11b} [because $X_3(\theta,1,\xi) = X_1(\theta ; 0) = \text{constant}$].

The Euler potentials for this stationary and axisymmetric configuration fall in the category \eqref{genpsi} where in Poincar\'e coordinates
\begin{equation}\label{eqpsi11cc}
\boxed{
\psi(r, \theta)= \Biggl(\frac{c_1 X_{3}}{r}\Biggr)^{h}, \qquad
I(\psi)=\mp hc_{1} \sqrt{\xi} \psi^{\frac{h-1}{h}}, \qquad
\Omega(\psi)= c_1 \frac{h-1}{h} \psi^{-\frac{1}{h}}}
\end{equation}

Since $2\pi \psi(r,\theta)$ is the magnetic flux through the loop of revolution defined by $(r,\theta)$, the requirement of having no singular magnetic flux at the north and south poles is equivalent to the boundary conditions $X_3(0) =X_3(\pi) = 0$. 

These potentials allow us to recognize the solution as the one described in \cite{Zhang:2014pla} upon identifying their quantities in terms of ours as 
\bea \nonumber
\alpha = -h, \quad f(\theta) = (c_1 X_3)^{h}, \quad g(\theta)= \frac{h-1}{h}\frac{1}{X_3}, \quad C=c_1\frac{h-1}{h}, \quad D=\pm \frac{c_1}{2\pi}\sqrt{\xi}.
\eea

\subsubsection*{Poincar\'e magnetic}
\begin{empheq}[box=\fbox]{align*}
A_{(h, 0)}&=c_2^h \Phi^{h}\Bigl[ X_{4}\mu^{2}\pm X_{4}^{\frac{h-1}{h}}\mu^{3}\Bigr]\\
F_{(h,0)}&=c_2^h \Phi^{h-1}\Biggl[\pm h X^{\frac{h-1}{h}}_{4}w^{1} + k \gamma X'_{4}w^{2} - \frac{1}{k\gamma}X'_{4}w^{3} + hX_{4}(w^{4}-w^{5})\Biggr]\\
J_{(h, 0)}&=c_2^h \frac{\Phi^{h}}{\Gamma} \Biggl[h(h-1)X_{4}^{\frac{h-2}{h}}\hat{Q}_0 - k X_{4} \frac{C(\theta ;h)}{\gamma(k^2\gamma^2-1)}\frac{\Phi H_{+}}{\sqrt{2}\Gamma}  \mp h(h-1)X_{4}^{\frac{h-1}{h}}\mu^{3}\\
&\qquad \pm\frac{h(h-1)}{\gamma}X_{4}^{-\frac{1}{h}}X'_4\mu^{4}\Biggr]
\end{empheq}
where $C(\theta; h)$ is given by
\begin{align}
C(\theta; h)=h \gamma \left((h-1) X_{4}^{-2/h}-1\right)-\frac{2 \gamma' X_{4}'}{X_{4}}+h k^2 \gamma^3.
\end{align}

Since $\p_t F = 0$, the Euler potentials for this stationary and axisymmetric configuration are given by \eqref{genpsi} where in Poincar\'e coordinates
\begin{equation}\label{eqpsi112}
\boxed{
\psi(r, \theta)=\Biggl(\frac{c_2}{r}\Biggr)^{h}X_{4}, \qquad
I(\psi)=\mp hc_{2}\psi^{\frac{h-1}{h}}, \qquad
\Omega(\psi)=0}
\end{equation}

Since $2\pi \psi(r,\theta)$ is the magnetic flux through the loop of revolution defined by $(r,\theta)$, the requirement of having no singular magnetic flux at the north and south poles is equivalent to the boundary conditions $X_4(0) =X_4(\pi) = 0$. In turn, regularity of the current at the poles then requires $h \geq 2$.

\subsubsection*{Poincar\'e magnetic}

\begin{empheq}[box=\fbox]{align*}
A_{(h, 0)}&=  \Phi^{h} X_{1}\mu^{2} \\
F_{(h,0)}&=\Phi^{h-1}\Bigl[ k\gamma X_{1}^{' } w^{2} - \frac{1}{k \gamma} X_{1}^{' } w^{3} + h X_{1} (w^{4}-w^{5})\Bigr]\\
J_{(h, 0)}&=\Phi^{h-1}\frac{2 k \gamma' X'_1 +h k\gamma X_1 (1-k^2 \gamma^2)}{(-1+k^2\gamma^2) \gamma \Gamma} \frac{\Phi H_{+}}{\sqrt{2}\Gamma}
\end{empheq}
where $X_1=X_1(\theta ; \Delta(h))$. Currents with different values of $h$ are collinear so one might linearly superpose such solutions. 

The Euler potentials for this stationary and axisymmetric configuration are
\begin{equation}\label{eqpsi11}
\boxed{
\psi(r, \theta)=\frac{X_{1}}{r^h} , \qquad
I(\psi)=0, \qquad
\Omega(\psi)=0}
\end{equation}

Since $2\pi \psi(r,\theta)$ is the magnetic flux through the loop of revolution defined by $(r,\theta)$, the requirement of having no singular magnetic flux at the north and south poles is equivalent to the boundary conditions $X_1(0) =X_1(\pi) = 0$.

\subsubsection*{Poincar\'e nontoroidal}

\begin{empheq}[box=\fbox]{align*}
A_{(h, 0)}&=\Phi^{h}X_{2}\Bigl[h \mu^{1} \pm\sqrt{c_{1}}\mu^{3}\Bigr]\\
F_{(h,0)}&=h\Phi^{h-1}X_{2}\Biggl(\pm \sqrt{c_{1}}w^{1} - \gamma \frac{X'_{2}}{X_{2}}w^{2}-\frac{(h-1)}{k}w^{4}\Biggr)\\
J_{(h, 0)}&=\Phi^{h}\frac{hX_{2}}{\gamma^{2}\Gamma}\Biggl[k\gamma^{2}[(h-1)\gamma^{2}-c_{1}]\hat{Q}_0 + c_{1}\frac{\Phi H_{+}}{\sqrt{2}\Gamma} \mp (h-1)\gamma^{2}\sqrt{c_{1}}\mu^{3}\\
&\qquad \pm\gamma \sqrt{c_{1}}\frac{X'_{2}}{X_{2}}\mu^{4}\Biggr]
\end{empheq}
where $X_2=X_2(\theta; h, c_1)$. 

For $h=0$ this solution has vanishing field strength so is pure gauge. For $c_{1}=0$ we get a Poincar\'e electric and nontoroidal solution, which has the special property to admit descendants solutions. The solution is real in Poincar\'e coordinates for $c_1 \geq 0$, $h \in \mathbb R$. 

Since $i_\phi F = 0$, the electromagnetic field in terms of Euler potentials takes the special form \eqref{Fs1} where  in Poincar\'e coordinates,
\begin{equation}
\boxed{
\chi(r, \theta)=h \frac{X_2}{r^{h-1}}, \qquad I(\chi)= \mp\sqrt{c_1} \chi}\label{eq:r1}
\end{equation}
We observe that in order to prevent singular line currents we need to enforce a vanishing polar current $I$ at north and south poles, which requires the existence of the boundary conditions $X_2(0)=X_2(\pi)=0$ for $c_1 \neq 0$. For $c_1 = 0$ the polar current vanishes but the electrostatic potential $\chi$ is constant on the north and south poles.

\subsection*{$(h = 0, q \neq 0)$-eigenstates}
\subsubsection*{Poincar\'e electric}

\begin{empheq}[box=\fbox]{align*}
A_{(0, q)}&=\lambda^{q}e^{\pm\int \frac{q}{\gamma}d\theta}\mu^{1}, \\
F_{(0, q)}&=\lambda^{q}e^{\pm\int \frac{q}{\gamma}d\theta}\Bigl(\mp q w^{1} +\frac{1}{k}w^{4}+i q w^{6}\Bigr)\\
J_{(0, q)}&=\lambda^q\frac{k}{\Gamma}e^{\pm\int \frac{q}{\gamma}d\theta}\Biggl[-(q^2+\gamma^2)\hat{Q}_0 + {i}q(\pm q \mu^3-\mu^4)\Biggr]
\end{empheq}
The field strength is singular at either the north or south pole depending upon the sign. The solution might however be interesting if it is split at the equator with regular north and south branches. 

In terms of Euler potentials and in Poincar\'e coordinates, we are in the case \eqref{EP1} where 
\begin{equation}
\boxed{
\chi_1(r,\theta)=r e^{\pm\int \frac{q}{\gamma}d\theta}, \qquad \chi_2(r,\theta)=0,\qquad \kappa(\phi_1)=0}
\end{equation}
The vanishing of $\chi_2$ is related to the absence of magnetic field.

\subsection*{$(h = 1, q \neq 0)$-eigenstates}

\subsubsection*{Poincar\'e electric - admitting descendants}

\begin{empheq}[box=\fbox]{align*}
A_{(1, q)}&=\Phi \lambda^{q} e^{\pm\int \frac{q}{\gamma}d\theta}\mu^{1}\\
F_{(1,q)}&=q \lambda^{q} e^{\pm\int \frac{q}{\gamma}d\theta}(\mp  w^{2}+i w^{6})\\
J_{(1,q)}&=\Phi \lambda^{q} e^{\pm\int \frac{q}{\gamma}d\theta}\frac{kq^{2}}{\Gamma}(-Q_{0}\pm i \mu^{3})
\end{empheq}
The field strength is again singular at either the north or south pole depending upon the sign. The solution might however be interesting if it is split at the equator with regular north and south branches. The direction of the current does not depend upon $q$ and therefore one can linearly superpose solutions with different $q$'s. The solution admits descendants. 

In terms of Euler potentials and in Poincar\'e coordinates, we are again in the case \eqref{EP1} where 
\begin{equation}
\boxed{
\chi_1(\theta)=e^{\pm\int \frac{q}{\gamma}d\theta}, \qquad \chi_2(r,\theta)=0, \qquad \kappa(\phi_1)=0}
\end{equation}

\subsection*{$(h(q)=\pm {i}kq,q \neq 0$)-eigenstates}
\noindent The two following classes of solutions feature a charge-dependent weight $h$.

\subsubsection*{Poincar\'e generic}

\begin{empheq}[box=\fbox]{align*}
A_{(h(q), q)}&=\Phi^{h}\lambda^{q}e^{s_2 \int\frac{d\theta}{\gamma}}\Bigl({i}kq\mu^{1}+{i}q \mu^{2}+s_2 \mu^3\Bigr),\qquad s_2 = -1 \text{ or } 1\\
F_{(h(q), q)}&= i k q \Phi^{h-1} \lambda^{q}e^{s_2 \int\frac{d\theta}{\gamma}}\Bigl(\pm s_2 w^{1} -s_2 w^{2} + \frac{1}{k}w^{4} \mp iq w^{5} + iq w^{6} \Bigr)\\
J_{(h(q), q)}&=\Phi^{h}\lambda^{q}\frac{kq}{\Gamma}e^{s_2 \int\frac{d\theta}{\gamma}}\Biggl([ \mp q + i k(q^{2}-1) -k(i \pm kq)\gamma^{2}]{\hat Q}_{0} \\
&\qquad + [ \pm kq - \frac{ i(q^{2}-1)}{\gamma^2}]\Bigl(\frac{\Phi H_{+}}{\sqrt{2}\Gamma} \pm \mu^{4}\Bigr) \pm i s_2 \mu^{3}\Biggr)
\end{empheq}
The solution is pure gauge when $q=0$. 
In terms of Euler potentials and in Poincar\'e coordinates, we are in the case \eqref{EP1} where 
\begin{equation}
\boxed{
\chi_1(r,\theta)=\frac{{i}kq}{r^{h-1}}e^{s_2 \int\frac{d\theta}{\gamma}}, \qquad \chi_2(r)=\mp\frac{1}{r},\qquad \kappa(\phi_1) = 0}
\end{equation}
	

\subsubsection*{Poincar\'e generic}

\begin{empheq}[box=\fbox]{align*}
A_{(h(q), q)}&=\Phi^{h}\lambda^{q}\Bigl[k\mu^{1}+\mu^{2}\Bigr]\\
F_{(h(q), q)}&= \Phi^{h-1} \lambda^{q}\Bigl(w^{4} + ikq (\mp w^{5} +w^{6}) \Bigr)\\
J_{(h(q), q)}&=\Phi^{h}\lambda^{q}\frac{k}{\Gamma}(\pm q + ik\gamma^{2})\Biggl( (i \pm kq)Q_{0} +\frac{q}{\gamma^{2}}\Bigl(\mp \frac{\Phi H_{+}}{\sqrt{2}\Gamma}-\mu^{4}\Bigr)\Biggr)
\end{empheq}

In terms of Euler potentials and in Poincar\'e coordinates, we are in the case \eqref{EP1} where 
\begin{equation}
\boxed{
\chi_1(r)=\frac{k}{r^{h-1}}, \qquad \chi_2(r)=\pm\frac{1}{r},\qquad \kappa(\phi_1)=0}
\end{equation}

\subsection*{$ (h(q)=1 \pm {i}kq,q \neq 0$)-eigenstates}

\subsubsection*{Poincar\'e generic - null}

\begin{empheq}[box=\fbox]{align*}
A_{(h(q), q)}&=\Phi^{h}\lambda^{q}\Bigl[ha_1(\theta)\mu^{1}\pm {i}qa_1(\theta)\mu^{2}\pm\gamma a_1'(\theta)\mu^3\Bigr]\\
F_{(h(q), q)}&=h \Phi^{h-1}\lambda^{q}\Bigl[\gamma a'_{1}(\theta)(\pm w^{1} - w^{2}) + iqa_{1}(\theta)(\mp w^{5} + w^{6})\Bigr]\\
J_{(h(q), q)}&=\Phi^{h}\lambda^{q}\frac{(1\pm{i}kq)[q^{2}a_{1}-\gamma\partial_{\theta}(\gamma a'_{1})]}{\gamma^{2}\Gamma}\Bigl(k\gamma^{2}\hat{Q}_0 - \frac{\Phi H_{+}}{\sqrt{2}\Gamma} \mp \mu^{4}\Bigr)
\end{empheq}
where $a_1(\theta)$ is an arbitrary function. It is a null solution ($F_{\mu\nu}F^{\mu\nu}  = 0$). 
The current is nonvanishing for $q \neq 0$. Indeed, the current vanishes when $a_{1}(\theta)$ obeys $a''_1+\frac{\gamma'}{\gamma}a'_1-\frac{q^2}{\gamma^2}a_1=0$. After a closer look at this differential equation, we conclude that a solution is given by $X_2(\theta; \Delta(h(q), q)=0,c_1=q^2)$. The constraint $\Delta(h(q), q)=0$ however implies $q=0$ in contradiction to our assumption $q\neq0$. 

In terms of Euler potentials and in Poincar\'e coordinates, we are in the case \eqref{EP1} where
\begin{equation}
\boxed{
\chi_1(r,\theta)=\frac{ha(\theta)}{r^{h-1}}, \qquad \chi_2(r)=\pm\frac{1}{r},\qquad \kappa(\phi_1) = 0}
\end{equation}

\subsection*{$ (h=1 ,q = 0$)-eigenstates}

\subsubsection*{Poincar\'e nontoroidal - null}

\begin{empheq}[box=\fbox]{align*}
A_{(1, 0)}&=\Phi \Bigl[a_{1}(\theta)\mu^{1}\pm \sqrt{c_{3}+[\gamma a_{1}'(\theta)]^{2}}\mu^{3}\Bigr]\\
F_{(1, 0)}&=\pm \sqrt{c_{3}+[\gamma a_{1}'(\theta)]^{2}} w^{1} - \gamma a_1'(\theta) w^{2}\\
J_{(1, 0)}&=\Phi \frac{\partial_{\theta}(\gamma a'_{1})}{\Gamma}\Biggl[-k \gamma \hat{Q}_0 + \frac{\Phi H_{+}}{\sqrt{2}\Gamma} \pm \frac{a'_{1}}{\sqrt{c_{3}+[\gamma a_{1}'(\theta)]^{2}}}\mu^{4}\Biggr]
\end{empheq}
It is a null solution ($F_{\mu\nu}F^{\mu\nu} = 0$). This class of solutions does not overlap with the class above. 
The current is vanishing when $\partial_{\theta}(\gamma a'_{1})=0$, i.e., when $a_1=c_2+\int \frac{c_1}{\gamma}d\theta$, where $c_1$ and $c_2$ are two real constants. Regularity fixes $c_1=0$ so only constant $a_1$ solutions obey Maxwell's equations. 

In terms of Euler potentials, the field strength reads as \eqref{Fs1} where 
\begin{equation}
\boxed{
\chi(r, \theta)=a_{1}(\theta), \qquad I(\chi)= \mp \sqrt{c_{3}+[\gamma \chi']^{2}}}
\end{equation}
We see that we need $c_3 = 0$ in order to have a regular configuration (no polar current on the $\theta = 0$ axis). We also impose that $a_1(\theta)$ must be regular at the poles. 

\subsection*{$(h=0 ,q = 0$)-eigenstates}

\subsubsection*{Poincar\'e generic - admitting descendants}

\begin{empheq}[box=\fbox]{align*}
A_{(0,0)} &= \Bigl(c_1 +c_3 \int \frac{d\theta}{\gamma} \Bigr)\mu^1 \\
F_{(0,0)}&= \Phi^{-1}\Bigl( -c_{3}w^{2} + \frac{c_1 +c_3 \int \frac{d\theta}{\gamma}}{k} w^{4} \Bigr)\\
J_{(0,0)} &= -\frac{k\gamma^{2}}{\Gamma}\Bigl(c_1 +c_3 \int \frac{d\theta}{\gamma}\Bigr)Q_{0}
\end{empheq}
The solution with $c_3 \neq 0 $ is singular at the poles. Indeed, 
\bea
\int \frac{d\theta}{\gamma} = \frac{\cos(\theta)}{2} - \ln\Bigl[\tan\Bigl(\frac{\theta}{2}\Bigr)\Bigr].
\eea
Therefore we fix $c_3 = 0$. The solution then becomes electric without toroidal fields. In fact, it is just the maximally symmetric solution. It is related to \eqref{eq: 19b} by a gauge transformation.


\providecommand{\href}[2]{#2}\begingroup\raggedright\endgroup

\end{document}